\documentclass[prd,a4paper,showpacs,showkeys,preprintnumbers,twocolumn,longbibliography,nofootinbib]{revtex4-1}

\usepackage{amsmath}
\usepackage{amssymb}
\usepackage[hidelinks]{hyperref}
\usepackage{times}
\usepackage{dsfont}
\usepackage{commath}
\usepackage{xfrac}
\renewcommand{\vec}[1]{\boldsymbol{\mathbf{#1}}}
\newcommand{\GF}{G_{\rm F}}
\newcommand{\sm}{\scalebox{2}[1]{-}}
\newcommand{\smi}{\scalebox{0.9}[0.9]{-}}
\newcommand{\spi}{\scalebox{0.75}[0.75]{+}}

\begin{document}
\preprint{MPP-2015-45}

\title{Neutrino propagation in media: Flavor, helicity, and pair correlations}

\author{A. Kartavtsev}
\email[\,]{alexander.kartavtsev@mpp.mpg.de}

\author{G. Raffelt}
\email[\,]{raffelt@mpp.mpg.de}

\author{H. Vogel}
\email[\,]{hvogel@mpp.mpg.de}

\affiliation{Max-Planck-Institut f\"ur Physik, F\"ohringer Ring 6, 80805 M\"unchen, Germany}

\begin{abstract}
Neutrinos propagating in media (matter and electromagnetic fields)
undergo flavor and helicity oscillations, where helicity transitions are
instigated both by electromagnetic fields and matter currents. In addition,
it has been shown that correlations between neutrinos and antineutrinos of
opposite momentum can build up in anisotropic media. We rederive the
neutrino equations of motion in the mean-field approximation for
homogeneous yet anisotropic media, confirming previous results except for a
small correction in the Majorana case. Furthermore, we derive the
mean-field Hamiltonian induced by neutrino electromagnetic interactions.
We also provide a phenomenological discussion of pair correlations in comparison
with helicity correlations. 
\end{abstract}

\maketitle

\section{\label{sec:introduction}Introduction}

Neutrino flavor conversion in vacuum \cite{Pontecorvo:1967fh, Gribov:1968kq},
in matter \hbox{\cite{Wolfenstein:1977ue, Wolfenstein:1979ni, Mikheev:1986gs,
Mikheev:1986if, Kuo:1989qe, Dighe:1999bi}},
or self-induced flavor conversion in a gas of interacting neutrinos
\cite{Pantaleone:1992eq, Samuel:1993uw, Pantaleone:1994ns, Sawyer:2005jk, Duan:2006an,
 Hannestad:2006nj, Balantekin:2006tg, Dasgupta:2009mg, Duan:2010bg, Friedland:2010sc,
 Banerjee:2011fj, Galais:2011gh, Pehlivan:2011hp, Raffelt:2011yb, Cherry:2012zw,
 deGouvea:2012hg, Raffelt:2013rqa, Duan:2014gfa, Mirizzi:2015fva}
provide a rich phenomenology of very practical experimental and
astrophysical importance. The data leave no room for doubt that
neutrinos have small but nonvanishing masses. One consequence is that
neutrinos have small electromagnetic dipole and transition moments
\cite{Giunti:2014ixa}. These lead to spin and spin-flavor
oscillations in strong electromagnetic fields
\cite{Werntz:1970, Cisneros:1970nq, Lim:1987tk, Akhmedov:1987nc}.
Actually, polarized matter or
matter currents alone instigate spin and spin-flavor transitions of
massive neutrinos, having effects similar to electromagnetic fields
\hbox{\cite{Egorov:1999ah, Grigoriev:2002zr, Studenikin:2004bu, Vlasenko:2013fja, Cirigliano:2014aoa}}.

For neutrinos streaming from a supernova core, the background medium
may contain currents. Moreover, the neutrino stream itself 
provides an unavoidable nonisotropic background. In addition,
self-induced flavor conversion in an interacting neutrino gas requires
unstable modes in flavor space (run-away solutions). If such solutions
exist, even small perturbations or otherwise small effects can grow
exponentially. In this sense, it is never obvious if a seemingly small
effect can get amplified by an instability to play an important role
after all.  Therefore, it is interesting to study if an interacting
neutrino gas can amplify helicity conversion effects
\cite{Vlasenko:2014bva} which otherwise are very small.

Flavor oscillations lead to correlations building up between
neutrinos of different flavor. If $a^\dagger_\alpha$ is the creation
operator of a neutrino in flavor state $\alpha$ with a certain
momentum $\vec{p}$, the initially prepared system can be described by the
occupation number $\langle a^\dagger_\alpha a_\alpha\rangle$. One way
of looking at flavor oscillations is that ``flavor off-diagonal''
occupation numbers of the type $\langle a^\dagger_\alpha
a_\beta\rangle$ develop and oscillate
\cite{Dolgov:1980cq, Rudzsky:1990, Sigl:1992fn}. One unifies these
expressions in a density matrix $\rho$ with components
\smash{$\rho_{\alpha\beta}=\langle a^\dagger_\beta a_\alpha\rangle$}. It
evolves according to the commutator equation $i\dot\rho=[{\sf
    H},\rho]$, where ${\sf H}$ is the Hamiltonian matrix, consisting of
oscillation frequencies. For vacuum oscillations we have ${\sf H}={\sf
  M}^2/2E$, where ${\sf M}^2$ is a matrix of squared neutrino
masses. Similar descriptions pertain to spin and spin-flavor
oscillation, where  the indices now indicate various states
of spin and/or flavor.

It was recently stressed that yet another form of correlations,
hitherto neglected in the context of neutrino propagation, can build
up in nonisotropic media \cite{Volpe:2013uxl, Vaananen:2013qja, Serreau:2014cfa}.
If $a^\dagger_{\vec{p}}$ is the creation operator of a
massless neutrino in mode $\vec{p}$ and \smash{$b^\dagger_{-\vec{p}}$} the one for an
antineutrino with opposite momentum, correlators of the form
\smash{$\kappa_{\vec{p}}=\langle b_{-\vec{p}}a_{\vec{p}}\rangle$} and
\smash{$\kappa^\dagger_{\vec{p}}=\langle a^\dagger_{\vec{p}}b^\dagger_{-\vec{p}}\rangle$}
will build up, the latter corresponding to the creation of a
particle-antiparticle pair with vanishing total momentum. Because
massless neutrinos and antineutrinos have opposite
helicity, this pair has total spin~1 so that its creation requires a
medium current transverse to $\vec{p}$ to satisfy angular-momentum
conservation. This requirement is analogous to the case of helicity
transitions where we also need a transverse current or magnetic field
for the same reason.  Including flavor and spin degrees of freedom
expands the ``pair correlations'' $\kappa$ and $\kappa^\dagger$ to
become matrices similar to $\rho$.

To develop more intuition about the meaning of the pair correlations,
we consider a single mode $\vec{p}$ of neutrinos
and $-\vec{p}$ of antineutrinos. We define \smash{$\rho_{\vec{p}}=\langle
a^\dagger_{\vec{p}} a_{\vec{p}}\rangle$} and for antineutrinos
\smash{$\bar\rho_{\vec{p}}=\langle b^\dagger_{-\vec{p}} b_{-\vec{p}}\rangle$}
involving the opposite momentum. Following the earlier literature
\cite{Volpe:2013uxl, Vaananen:2013qja, Serreau:2014cfa}, we unify
these expressions in an extended density matrix
\begin{equation}\label{eq:simple-R-equation}
{\sf R}=\begin{pmatrix}\rho&\kappa\cr\kappa^\dagger&1-\bar\rho\cr\end{pmatrix}
=\begin{pmatrix}
\langle a^\dagger_{\vec{p}} a_{\vec{p}}\rangle&\langle b_{-\vec{p}}a_{\vec{p}}\rangle\cr
\langle a^\dagger_{\vec{p}}b^\dagger_{-\vec{p}}\rangle&\langle b_{-\vec{p}} b^\dagger_{-\vec{p}}\rangle\cr
\end{pmatrix}\,,
\end{equation}
which obeys an equation of motion of the form
\cite{Volpe:2013uxl, Vaananen:2013qja, Serreau:2014cfa}
\begin{equation}\label{eq:EOM1}
i\dot{\sf R}=[{\sf H},{\sf R}]\,.
\end{equation}
If the background is a current
moving in the transverse direction with velocity $\beta$, the
Hamiltonian matrix is found to be
\begin{equation}\label{eq:simplehamiltonian}
{\sf H}=E\,\begin{pmatrix}1&0\cr 0&-1\cr\end{pmatrix}
+V\,\begin{pmatrix}1&-\beta\cr-\beta&1\cr\end{pmatrix}\,,
\end{equation}
where $E=|\vec{p}|$. For $\nu_\mu$ or $\nu_\tau$ neutrinos the usual matter
potential is \smash{$V=\GF n_n/\sqrt{2}$}, where $n_n$ is the neutron density.

This commutator equation has the same structure that one
encounters for the evolution of any two-level system and in
particular for two-flavor or helicity oscillations,
of course with a different matrix ${\sf H}$ for each
case. However, what specifically are the two states that are being
mixed by the matter current in the pair-correlation case?

The answer
becomes evident if one considers the evolution of states
rather than correlators. Our simple
system is described by the four basis states $|00\rangle$,
$|10\rangle$, $|01\rangle$ and $|11\rangle$, where the first entry
refers to $\nu(\vec{p})$ and the second to $\bar\nu(-\vec{p})$. A homogeneous
background medium cannot mix states of different total momentum, so the
single-particle states must evolve independently as $i\partial_t|10\rangle=
(E+V)|10\rangle$ and  $i\partial_t|01\rangle=
(E-V)|01\rangle$, i.e., they simply suffer the usual energy shift by
the weak potential of the medium. This leaves us
with $|00\rangle$ and $|11\rangle$ which both have zero momentum and
therefore can be mixed by a homogeneous medium.
The former has spin~0, the latter spin~1, so for the medium to mix
them, it must provide a transverse vector in the form of a current or
a spin polarization. If $A_{00}$ and $A_{11}$ are the
amplitudes of $|00\rangle$ and $|11\rangle$, respectively, we will
show later that Eq.~(\ref{eq:EOM1}) corresponds to
\begin{equation}\label{eq:0011}
i\partial_t\begin{pmatrix}A_{00}\cr A_{11}\cr\end{pmatrix}
=\begin{pmatrix}0&\beta V\cr\beta V&2E\cr\end{pmatrix}
\begin{pmatrix}A_{00}\cr A_{11}\cr\end{pmatrix}\,.
\end{equation}
Therefore, it is the empty and the completely filled states that are
being mixed and that oscillate. The true ground state
of our system is not $|00\rangle$, but a suitable combination of $|00\rangle$
and  $|11\rangle$ which follows from diagonalizing the matrix in Eq.~(\ref{eq:0011}).

As we have noted, any two-level system is equivalent to an abstract
spin-$\frac{1}{2}$ system. In two-flavor oscillations, the ``spin''
represents the two flavor states. In the pair-correlation case, ``spin
up'' means ``empty'' and ``spin down'' means ``full with a pair.''
This interpretation is analogous to Anderson's ``pseudo spin'' devised
to describe Cooper pairs in the context of superconductivity
\cite{Anderson:1958zza}. A coherent superposition of these two spin
states, represented in our case by the pair correlations, corresponds
to a coherent superposition of $|00\rangle$ and $|11\rangle$.

In analogy to the example of superconductivity,
another way to think about these phenomena is in terms of Bogolyubov
transformations of the creation and annihilation operators. If we think of a
single momentum mode $\vec{p}$ of mixed neutrinos in vacuum, the operators $a_{\nu_e}$
and $a_{\nu_\mu}$ in the flavor basis are rotated by a unitary transformation with
mixing angle $\vartheta$ to form new operators $c_\vartheta a_{\nu_e} + s_\vartheta a_{\nu_\mu}$
and  $c_\vartheta a_{\nu_e}-s_\vartheta a_{\nu_\mu}$, and similarly for the creation operators,
to form a new set of canonically anticommuting operators, now describing
neutrinos in the mass basis. Describing flavor oscillations in terms of
time-dependent Bogolyubov transformations can be especially illuminating to
understand quantum statistics in mixing phenomena for both bosons and fermions
\cite{Raffelt:1991ck}. Pair correlations correspond to the
same idea where the mixing is between $a_{\vec{p}}$ and \smash{$b^\dagger_{-\vec{p}}$} with a
mixing angle corresponding to the unitary transformation that diagonalizes
the matrix in Eq.~(\ref{eq:0011}). The state $|00\rangle$ defined in
the Bogolyubov-transformed basis is the ground state of
the system and no longer oscillates into the new $|11\rangle$ state.

The goal of our paper is two-pronged. On the one hand we reconsider
the mean-field equations of motion for massive neutrinos propagating
in a background medium that can consist of matter and neutrinos, and
that is homogeneous but not isotropic. Besides the usual flavor
oscillations in matter, the resulting phenomena include spin and
spin-flavor oscillations as well as pair correlations.

As a second goal, we provide a phenomenological discussion of the
interpretation of the pair correlations in the context of neutrino
oscillation problems in dense media. Ultimately, our community needs
to develop an understanding if, from a practical perspective, we need
to worry about pair correlations and helicity oscillations in the
supernova context.

The supernova environment is characterized by
small neutrino energies of at most some 200~MeV (for degenerate
$\nu_e$), i.e., small compared to $W$ and $Z$ masses so that it
suffices to describe neutrino interactions in terms of an effective
current-current Hamiltonian. In the early Universe, where the chemical
potentials of background particles are small, one has to worry about
corrections from the electroweak gauge-boson propagators even at low
temperatures \cite{Notzold:1987ik}. The supernova environment, in
contrast, has large densities of background particles and this concern
is moot.

On the mean-field level, the current of background particles is a
classical quantity. For example, the neutral-current interaction of a
neutrino with neutrons is given by the Hamiltonian density ${\cal
  H}=\sqrt{2}\GF\,[\bar\nu\gamma_\mu P_L\nu]\,I_n^\mu$, where $\GF$ is
the Fermi constant, $\nu$ is the neutrino Dirac field, $P_L$ is the
left-handed projector, and $I_n^\mu$ is the neutron current.  If the
current is homogeneous, $H=\int d^3{\bf x}\,{\cal H}$ is effectively a
``forward'' Hamiltonian: it couples, e.g.,
\smash{$a^\dagger$} and $a$ of equal momenta.  Following the previous
literature \cite{Rudzsky:1990, Sigl:1992fn, Serreau:2014cfa}, the
evolution of, e.g., the annihilator for a neutrino of
mass eigenstate $i$ in mode $\vec{p}$
is given by the Heisenberg equation of motion $i\partial_t
a_i(t,\vec{p})=[a_i(t,\vec{p}),H]$.  It is then straightforward to
find the equations of motion of bilinears of the form
\smash{$a_i^\dagger(t,\vec{p})a_j(t,\vec{p})$}, of their expectation value
\smash{$\langle a_i^\dagger(t,\vec{p})a_j(t,\vec{p})\rangle$}, of the entire
matrix $\rho$, and then of the extended matrix ${\sf R}$ which also
includes pair correlations.

It is largely a
cumbersome bookkeeping exercise to obtain, for neutrinos with mass,
all the components of the Hamiltonian matrix ${\sf H}$
appearing in Eq.~(\ref{eq:EOM1}) when ${\sf R}$ involves
all components of spin and flavor. We perform this task separately
for Dirac neutrinos in Sec.~\ref{sec:dirac}, for Majorana neutrinos in Sec.~\ref{sec:majorana}, and
for Weyl neutrinos (massless two-component case) in Sec.~\ref{sec:weyl}.
These derivations closely parallel the recent paper by Serreau and Volpe
\cite{Serreau:2014cfa} and we will largely follow their notation to
avoid confusion. In the Majorana case, we find a small correction, but
otherwise our results agree.

The density matrix formalism
allows one to treat helicity oscillations induced by magnetic fields
and by matter currents  on equal footing for both Dirac and Majorana
fermions. We derive the  mean-field Hamiltonian induced by electromagnetic
fields in Sec.~\ref{sec:magnetic}. Concerning helicity oscillations
induced by matter currents, which we analyze in Sec.~\ref{sec:helicity}, our results
coincide with those of Volpe and Serreau, and parallel those of Vlasenko, Fuller, and
Cirigliano \cite{Vlasenko:2013fja, Cirigliano:2014aoa,
  Vlasenko:2014bva} as far as the mean-field limit is concerned. These
authors have derived the neutrino kinetic equations starting from
first field-theoretic principles and have carried the results beyond
the mean-field limit to include (nonforward) collision terms,
generalizing previous derivations \cite{Dolgov:1980cq, Rudzsky:1990,
  Sigl:1992fn}. We note in passing that one of their findings ---
helicity oscillations in a nonisotropic matter background --- had been
anticipated in several papers by Studenikin and collaborators who have
worked out the one-to-one correspondence to the effect of
electromagnetic fields \cite{Egorov:1999ah, Studenikin:2004bu}. Of course,
Vlasenko, Fuller, and Cirigliano also included
neutrino-neutrino interactions as an agent of helicity conversion and
carried their results beyond the mean-field limit.

Pair correlations have been studied in detail in condensed matter and 
nuclear physics, as well as in the context of pair creation in quantum field theory. 
On the other hand, in neutrino physics these concepts are less familiar. They
have been addressed only in a handful of papers in the context of leptogenesis, 
where  pair correlations have been studied from first principles
in a series of papers by Fidler, Herranen, Kainulainen and Rahkila
\cite{Herranen:2008di,Herranen:2008hi,Herranen:2009zi,Herranen:2011zg, Fidler:2011yq}.
In the context of neutrino propagation in supernovae, the only discussions so far 
appear in a series of papers 
by Volpe and collaborators \cite{Volpe:2013uxl,Vaananen:2013qja,Serreau:2014cfa}.
We address phenomenological aspects of pair correlations in 
Sec.~\ref{sec:partantipart} and compare them to helicity correlations.

Finally, in Sec.~\ref{sec:summary} we summarize the results and present
our conclusions.

\section{\label{sec:dirac}Dirac neutrino}

Our first goal is to derive the components of the Hamiltonian matrix
${\sf H}$ which governs the evolution equation (\ref{eq:EOM1}) for the
extended density matrix ${\sf R}$ including flavor, helicity,
and pair correlations. In this rather technical section, we begin
with the conceptually simplest case of three neutrino flavors which
are assumed to have Dirac masses. Therefore, helicity correlations
involve the sterile components of the neutrino field, which otherwise
are completely decoupled.

\subsection{Two-point correlators and kinetic equations}

In the simplest approximation, one can describe the state of a
neutrino gas in terms of one-particle distribution functions. They are
extended to include flavor and helicity coherence effects by promoting
the one-particle distribution functions to density matrices
\cite{Dolgov:1980cq, Rudzsky:1990, Sigl:1992fn, Vlasenko:2013fja,
  Serreau:2014cfa}. In terms of the usual creation and annihilation
operators, their components are
\begin{subequations}
\label{RhoDef}
\begin{align}
(2\pi)^3\delta(\vec{p}{-}\vec{k})\rho_{ij,sh} (t,\vec{p})&=
\langle a^\dagger_{j,h}(t,+\vec{k}) a_{i,s}(t,+\vec{p})\rangle\,,\\
(2\pi)^3\delta(\vec{p}{-}\vec{k})\bar{\rho}_{ij,sh} (t,\vec{p})&=
\langle b^\dagger_{i,s}(t,-\vec{p}) b_{j,h}(t,-\vec{k})\rangle\,,
\end{align}
\end{subequations}
where $i$ and $j$ are flavor indices in the mass basis, and $s$ and
$h\in\{+,-\}$ denote helicities. In this convention, the density matrix
for antineutrinos $\bar{\rho}_{ij,sh}(t,\vec{p})$ for momentum
${\vec{p}}$ actually corresponds to the occupation numbers of
antineutrinos with physical momentum $-\vec{p}$.

This convention is necessary to combine $\rho$ and $\bar\rho$ with the
pair correlations which are defined as \cite{Vaananen:2013qja,Serreau:2014cfa}
\begin{subequations}
\label{KappaDef}
\begin{align}
(2\pi)^3\delta(\vec{p}{-}\vec{k})\kappa_{ij,sh} (t,\vec{p})&=
\langle b_{j,h}(t,-\vec{k}) a_{i,s}(t,+\vec{p})\rangle\,,\\
(2\pi)^3\delta(\vec{p}{-}\vec{k})\kappa^\dagger_{ij,sh} (t,\vec{p})&=
\langle a^\dagger_{j,h}(t,+\vec{p}) b^\dagger_{i,s}(t,-\vec{k})\rangle\,,
\end{align}
\end{subequations}
and which involve opposite-momentum modes.

The kinetic equations for Eqs.~\eqref{RhoDef} and~\eqref{KappaDef} are
obtained with the Heisenberg equation of motion.  As we will show
below, in the mean-field approximation, and assuming spatial
homogeneity, the Hamiltonian of charged- and neutral-current
neutrino interactions can be written in the compact form
\begin{align}
\label{Heff}
H_{\rm mf}=\int d^3{\bf x}\,\bar{\nu}_i(t,\vec{x}) \Gamma_{ij}\nu_j(t,\vec{x})\,,
\end{align}
where summation over repeated indices is implied.
The kernel takes account of the background medium and is
\begin{align}
\Gamma_{ij} = \gamma_\mu P_L V^\mu_{ij}\,,
\end{align}
where $P_L=(1-\gamma_5)/2$ is the usual left-chiral projector.
The current of background matter $V^\mu_{ij}$ will be defined in Eq.~\eqref{SigmaDirac}.

The momentum-mode
decomposition of a Dirac neutrino field reads
\begin{align}\label{eq:Dirac-decomposition}
\nu_i(t,\vec{x}\,)=\int_{\vec{p},s}  e^{i\vec{p}\cdot\vec{x}}\nu_{i,s}(t,\vec{p})\,,
\end{align}
where \smash{$\int_{\vec{p},s}$} denotes the phase-space integration \smash{$\int
d^3\vec{p}/(2\pi)^3$} and the summation over helicities.
In the mass basis, the individual momentum modes are
\begin{align}
\label{nudecompositiondir}
\nu_{i,s}(t,\vec{p}) &= a_{i,s}(t,\vec{p})u_{i,s}(\vec{p})
+b^\dagger_{i,s}(t,-\vec{p})v_{i,s}(-\vec{p})\,.
\end{align}
The chiral spinors $u$ and $v$ are given in Appendix~\ref{sec:spinorproducts},
and the creation and annihilation operators satisfy the usual equal-time
anticommutation relation,
\begin{align}
\label{AnticommRels}
\{a_{i,s}(t,\vec{p}),a^\dagger_{j,h}(t,\vec{k})\}=(2\pi)^3 \delta(\vec{p}{-}\vec{k})\delta_{ij}\delta_{sh}\,.
\end{align}
Similar relations hold for the antiparticle operators $b$ and
$b^\dagger$.

As a next step, we contract the kernels $\Gamma_{ij}$ with the
spinors appearing in the mean-field Hamiltonian \eqref{Heff},
leading to the matrices \cite{Serreau:2014cfa}
\begin{subequations}
\label{GammasDef}
\begin{align}
\Gamma^{\nu\nu}_{ij,sh}(\vec{p})&\equiv\bar{u}_{i,s}(\vec{p})\Gamma_{ij}u_{j,h}(\vec{p})\,,\\
\Gamma^{\nu\bar\nu}_{ij,sh}(\vec{p})&\equiv\bar{u}_{i,s}(\vec{p})\Gamma_{ij}v_{j,h}(-\vec{p})\,,\\
\Gamma^{\bar\nu\nu}_{ij,sh}(\vec{p})&\equiv\bar{v}_{i,s}(-\vec{p})\Gamma_{ij}u_{j,h}(\vec{p})\,,\\
\Gamma^{\bar\nu\bar\nu}_{ij,sh}(\vec{p})&\equiv\bar{v}_{i,s}(-\vec{p})\Gamma_{ij}v_{j,h}(-\vec{p})\,,
\end{align}
\end{subequations}
in component form.  We can now bring Eq.~\eqref{Heff} to the desired
form bilinear in the creation and annihilation operators
\begin{align}
\label{HeffCrAn}
H_{\rm mf}=\int_{\vec{p}}\,\, \Bigl[& a^\dagger_{i,s}(\vec{p}) \Gamma^{\nu\nu}_{ij,sh}(\vec{p})
a_{j,h}(\vec{p})\nonumber\\[-2mm]
&{}+a^\dagger_{i,s}(\vec{p}) \Gamma^{\nu\bar\nu}_{ij,sh}(\vec{p})b^\dagger_{j,h}(-\vec{p})\nonumber\\
&{}+b_{i,s}(-\vec{p}) \Gamma^{\bar\nu\nu}_{ij,sh}(\vec{p}) a_{j,h}(\vec{p})\nonumber\\
&{}+b_{i,s}(-\vec{p}) \Gamma^{\bar\nu\bar\nu}_{ij,sh}(\vec{p})b^\dagger_{j,h}(-\vec{p})\Bigr]\,,
\end{align}
where we have omitted the time arguments to shorten the
notation. Summation over repeated indices is implied.

Using the Heisenberg equation of motion with this Hamiltonian 
one finds the extended equation of motion $i\dot{\sf R}=[{\sf H},{\sf R}]$,
see also Eq.\,\eqref{eq:EOM1}. The extended density matrix,
Eq.\,\eqref{eq:simple-R-equation}, and the Hamiltonian, Eq.\,\eqref{eq:simplehamiltonian},
generalize to \cite{Volpe:2013uxl}
\begin{align}
\label{RHstructure}
\sf{R}=
\begin{pmatrix}
\rho & \kappa\\
\kappa^\dagger & \mathds{1}-\bar{\rho}
\end{pmatrix}\quad {\rm and} \quad
\sf{H}=
\begin{pmatrix}
\sf{H}^{\nu\nu} & \sf{H}^{\nu\bar\nu}\\
\sf{H}^{\bar\nu\nu} & \sf{H}^{\bar\nu\bar\nu}
\end{pmatrix}\quad \,,
\end{align}
where the submatrices   ${\sf H}^{\nu\nu}=\Gamma^{\nu\nu}$, ${\sf H}^{\nu\bar\nu}=\Gamma^{\nu\bar\nu}$
etc.\,  and $\rho$,
$\kappa$, etc.\ are $6{\times}6$ matrices in helicity and flavor space. The product
between such matrices in the commutator is defined in the obvious way
\smash{$(A\cdot B)_{ij,sh}\equiv A_{in,sr} B_{nj,rh}$} with a summation over repeated indices.
In the following we write the matrix structure in the form of $2{\times}2$ matrices in helicity space,
\begin{equation}
\label{spinflavorstruct}
\begin{pmatrix}
\framebox{\hbox to 0pt{$\vphantom{+}$}$--$}_{~ij}&\framebox{$-+$}_{~ij}\\[6pt]
\framebox{$+-$}_{~ij}&\framebox{$++$}_{~ij}
\end{pmatrix}\,,
\end{equation}
where each entry is itself a $3{\times}3$ matrix in flavor space.

\subsection{Hamiltonian in the mean-field approximation}

After having established the overall structure of the kinetic equations 
we now turn to the interactions contributing to neutrino refraction in the
supernova environment. In this subsection we only consider charged- and
neutral-current neutrino interactions, whereas the analysis of the electromagnetic
interactions is postponed to Sec.\,\ref{sec:magnetic}.

\subsubsection{Charged-current interaction}

We begin with charged-current (cc) interactions with
background charged leptons. In the low-energy limit and after a Fierz
transformation, the usual current-current Hamiltonian density is
\begin{align}
\label{HCC}
\mathcal{H}^{\rm cc}&= \sqrt{2} \GF\sum\limits_{\alpha,\beta}
\bigl[\bar{\nu}_\alpha\gamma^\mu P_L\nu_\beta\bigr]
\bigl[\bar{\ell}_\beta\gamma_\mu (1-\gamma^5) \ell_\alpha\bigr]\,,
\end{align}
where $\alpha,\beta \in \{e,\mu,\tau\}$ are flavor indices.

To obtain the neutrino mean-field Hamiltonian  we
replace the second bracket by its expectation value. In the supernova
environment, the temperature is too low to support a substantial density of
muons or tauons, and we use only the electron background. Then we find in the
mass basis
\begin{align}
\label{HCCeff}
\mathcal{H}_{\rm mf}^{\rm cc}&= \sqrt{2}\GF\sum\limits_{i,j}
\bigl[\bar{\nu}_i\gamma^\mu P_L\nu_j\bigr]
\bigl[U^\dagger_{ie} I_{\rm cc}^\mu U_{ej}\bigr] \,,
\end{align}
where $U$ is the Pontecorvo-Maki-Nakagawa-Sakata (PMNS) matrix. We have
introduced a linear combination of vector and axial-vector charged electron
currents,
\begin{align}
\label{Jlept}
I_{\rm cc}^\mu\equiv  c_V\langle \bar{e} \gamma^\mu e \rangle
-c_A \langle \bar{e}\gamma^\mu \gamma^5 e \rangle\,,
\end{align}
where $c_V=c_A=1$. Because electrons are the only background particles
contributing to charged-current interactions and to simplify the notation, an ``$e$''
index is implied in $I_{\rm cc}^\mu$. If the electrons are not polarized, the
axial current vanishes and $I_{\rm cc}^\mu=J^\mu_{e}$, the ``convective''
electron current.

\subsubsection{Neutral-current interaction with matter}

The neutral-current (nc) interactions with matter are described in the mass
basis by the Hamiltonian density
\begin{align}
\label{HNC}
\mathcal{H}^{\rm nc}&= \sqrt{2}\GF\sum\limits_{i,f}\bigl[
\bar{\nu}_i\gamma_\mu P_L\nu_i\bigr]\bigl[\bar{\psi}_f\gamma^\mu (c_V^f{-}c_A^f\gamma^5)\psi_f]\,,
\end{align}
where $f$ denotes electrons, protons, and neutrons. The resulting
contribution to the mean-field Hamiltonian is
\begin{align}
\label{HNCeff}
\mathcal{H}^{\rm nc}_{\rm mf}&= \sqrt{2}\GF\sum\limits_{i}\bigl[
\bar{\nu}_i\gamma_\mu P_L\nu_i\bigr]\!\bigl[I_{\rm nc}^\mu+I_p^\mu+I_n^\mu],
\end{align}
where $I_{\rm nc}^\mu$ denotes the electron neutral current (index $e$
implied), whereas the other contributions refer to protons and neutrons as
explicitly indicated.

These currents are defined in analogy to Eq.~\eqref{Jlept} with the
appropriate coupling constants. For electrons, they are given by
$c_V=-\frac12+2\sin^2\theta_W$ (Weinberg angle $\theta_W$) and
$c_A=-\frac12$. For protons, \smash{$c_V=\frac12-2\sin^2\theta_W$}, i.e., the same as
for electrons with opposite sign, and for neutrons \smash{$c_V=-\frac12$}. For the
nucleon axial vector one often uses $c_A=\pm1.26/2$ in analogy to their
charged current. However, the strange-quark contribution to the nucleon spin
as well as modifications in a dense nuclear medium leave the exact values
somewhat open \cite{Raffelt:1993ix, Horowitz:2001xf}.

In an unpolarized and electrically neutral environment, the axial currents
disappear and the electron and proton contributions to the convective neutral
current cancel such that in Eq.~\eqref{HNCeff} we have $I_{\rm nc}^\mu+
I_p^\mu+I_n^\mu=-\frac12J_n^\mu$, where $J^\mu_n$ is the neutron convective
current. Neutrino refraction in such a medium depends only on the charged
electron current and the neutral neutron current.

\subsubsection{Neutrino-neutrino interaction}

The most complicated interaction is the neutral-current neutrino-neutrino
one. It is described in the mass basis by the Hamiltonian density
\begin{align}
\label{Hself}
\mathcal{H}^{\nu\nu}&= \frac1{\sqrt{2}}\GF\sum\limits_{ij}
\bigl[\bar{\nu}_i\gamma_\mu P_L\nu_i\bigr]\bigl[\bar{\nu}_j\gamma^\mu P_L\nu_j\bigr]\,.
\end{align}
To obtain the mean-field Hamiltonian bilinear in the neutrino fields we need
to replace products of two of the four neutrino fields in this expression by
their expectation value.

The only combinations that do not violate lepton number are of the type
$\langle\bar{\nu}_i \nu_j\rangle$ and $\langle\nu_i \bar{\nu}_j\rangle$,
where $i$ and $j$ can be equal or different. We denote the corresponding
mean field as
\begin{align}
\label{Inu}
I^\mu_{ij}\equiv \langle \bar{\nu}_j\gamma^\mu P_L\nu_i \rangle\,.
\end{align}
To simplify notation we avoid an explicit ``neutrino'' and ``nc'' index,
i.e., expressions of the type $I^\mu_{ij}$ always refer to the neutral neutrino current
for the mass states $i$ and $j$. An explicit expression in terms of the
density matrices and pair correlators will be given in
Eq.~\eqref{InuExplicit} below.

For the $i=j$ contractions it is sufficient to take the
expectation value of one of the square brackets in Eq.~\eqref{Hself}, leading
to the mean-field Hamiltonian
$\sqrt{2}\GF\sum_{ij} \bigl[\bar{\nu}_i\gamma_\mu P_L\nu_i\bigr]\,
I^\mu_{jj}$. For the $i\neq j$ contractions we use the Fierz identity to rewrite
the Hamiltonian as $[\bar{\nu}_i\gamma_\mu P_L\nu_j][\bar{\nu}_j\gamma^\mu P_L\nu_i]$ in
Eq.\,\eqref{Hself}, leading to the  contribution
$\sqrt{2}\GF\sum_{ij} \bigl[\bar{\nu}_i\gamma_\mu P_L\nu_j\bigr] I^\mu_{ij}$.
Altogether, we find
\begin{align}
\label{HselfEff}
{\cal H}^{\nu\nu}_{\rm mf}=
\sqrt{2}\GF \sum\limits_{ij}\bigl[\bar{\nu}_i\gamma_\mu P_L\nu_j\bigr]
\bigl[I_{ij}^\mu+\delta_{ij}\,\textstyle{\sum_k} I^\mu_{kk}\bigr]
\end{align}
for the neutrino-neutrino mean-field Hamiltonian.

\subsection{Components of the Hamiltonian matrix \boldmath{$\sf H$}}

Adding up Eqs.~\eqref{HCCeff}, \eqref{HNCeff}, and \eqref{HselfEff} we find
the overall mean-field current
\begin{align}
\label{SigmaDirac}
V^\mu_{ij} = \sqrt{2}\GF\bigl[&
U^\dagger_{ie} I_{\rm cc}^\mu  U_{ej}
+\delta_{ij}(I_{\rm nc}^\mu+I_p^\mu+I_n^\mu)\nonumber\\
&{}+I^\mu_{ij}+\delta_{ij}\,\textstyle{\sum_k} I^\mu_{kk}\bigr]\,.
\end{align}
The spinor contractions defined in Eq.~\eqref{GammasDef} lead to the
components of the Hamiltonian matrix ${\sf H}$ of the form
\begin{subequations}
\label{GammaDiracFull}
\begin{align}
\label{GammaDiracnunu}
{\sf H}^{\nu\nu}_{ij,sh}&=(\gamma_\mu P_L)^{\nu\nu}_{ij,sh}V^\mu_{ij}
+\delta_{sh}\delta_{ij}E_{i}\,,\\
{\sf H}^{\nu\bar\nu}_{ij,sh}&=(\gamma_\mu P_L)^{\nu\bar\nu}_{ij,sh}V^\mu_{ij}\,,\\
{\sf H}^{\bar\nu\nu}_{ij,sh}&=(\gamma_\mu P_L)^{\bar\nu\nu}_{ij,sh}V^\mu_{ij}\,,\\
{\sf H}^{\bar\nu\bar\nu}_{ij,sh}&=(\gamma_\mu P_L)^{\bar\nu\bar\nu}_{ij,sh}V^\mu_{ij}
-\delta_{sh}\delta_{ij}E_{i}\,,
\end{align}
\end{subequations}
where $E_i=(\vec{p}^2+m_i^2)^\frac12$ is the neutrino energy, and 
we have identified \smash{${\sf H}^{\nu\nu}=\Gamma^{\nu\nu}$}, 
\smash{${\sf H}^{\nu\bar\nu}=\Gamma^{\nu\bar\nu}$},
etc. We have used
the compact notation
\begin{subequations}\label{eq:contractiondefinition}
\begin{align}
(\gamma_\mu P_L)^{\nu\nu}_{ij,sh}        &\equiv\bar{u}_{i,s}(+\vec{p})\gamma_\mu P_L u_{j,h}(+\vec{p})\,,\\
(\gamma_\mu P_L)^{\nu\bar\nu}_{ij,sh}    &\equiv\bar{u}_{i,s}(+\vec{p})\gamma_\mu P_L v_{j,h}(-\vec{p})\,,\\
(\gamma_\mu P_L)^{\bar\nu\nu}_{ij,sh}    &\equiv\bar{v}_{i,s}(-\vec{p})\gamma_\mu P_L u_{j,h}(+\vec{p})\,,\\
(\gamma_\mu P_L)^{\bar\nu\bar\nu}_{ij,sh}&\equiv\bar{v}_{i,s}(-\vec{p})\gamma_\mu P_L v_{j,h}(-\vec{p})\,.
\end{align}
\end{subequations}
Later we will use similar expressions for contractions with other Dirac structures.
The neutrino mean-field current itself contains spinor contractions of this type and
can be expressed in terms of the density matrices and pair correlations as
\begin{align}
\label{InuExplicit}
I^\mu_{ij}=\int_{\vec{p},s,h} \Bigl[&(\gamma^\mu P_L)^{\nu\nu}_{ji,hs}\rho_{ij,sh}
\nonumber\\[-4pt]
&{}+(\gamma^\mu P_L)^{\nu\bar\nu}_{ji,hs}\kappa^\dagger_{ij,sh}\nonumber\\[2pt]
&{}+(\gamma^\mu P_L)^{\bar\nu\nu}_{ji,hs}\kappa_{ij,sh}
\nonumber\\
&{}+(\gamma^\mu P_L)^{\bar\nu\bar\nu}_{ji,hs}(\delta_{ij}\delta_{sh}-\bar{\rho}_{ij,sh})
\,\Bigr]\,.
\end{align}
Notice that in this case there is no implied summation over $i$ and $j$.
The fourth term contains a divergent vacuum contribution that must be renormalized.

We finally work out the spinor contractions explicitly to lowest order in neutrino
masses. To this end we introduce
\begin{align}
n^\mu=
\left(1,\hat{\vec p}\right)\,,\quad
\bar{n}^\mu=
\left(1,-\hat{\vec p}\right)\,,
\quad
\epsilon^\mu=
\left(0,\hat{\vec \epsilon}\right)\,,
\end{align}
where $\hat{\vec p}$ is a unit vector in the momentum direction
and the complex polarization vector $\hat{\vec \epsilon}$ spans
the plane orthogonal to~${\vec p}$ (see Appendix \ref{sec:spinorproducts} for more
details). We also use $\phi$ to denote the polar angle
of $\vec{p}$ in spherical coordinates. To lowest order in $m_i$, the
spinor contractions are then found to be
\begin{subequations}\label{Xinunu}
\begin{align}
(\gamma_\mu P_L)^{\nu\nu}_{ij,sh}  &\approx
\begin{pmatrix}
n_\mu & -e^{+i\phi}\frac{m_j}{2p}\epsilon^*_\mu \\
-e^{-i\phi}\frac{m_i}{2p}\epsilon_\mu & 0
\end{pmatrix}\,,\label{eq:contraction-nunu}\\[3pt]
(\gamma_\mu P_L)^{\nu\bar\nu}_{ij,sh}    &\approx
\begin{pmatrix}
-e^{+i\phi}\frac{m_j}{2p}n_\mu & \epsilon^*_\mu \\
0 & -e^{-i\phi}\frac{m_i}{2p}\bar{n}_\mu
\end{pmatrix}
\,,\\[3pt]
(\gamma_\mu P_L)^{\bar\nu\nu}_{ij,sh}    &\approx
\begin{pmatrix}
-e^{-i\phi}\frac{m_i}{2p}n_\mu & 0 \\[2pt]
\epsilon_\mu & -e^{+i\phi}\frac{m_j}{2p}\bar{n}_\mu
\end{pmatrix}
\,,\\[3pt]
(\gamma_\mu P_L)^{\bar\nu\bar\nu}_{ij,sh}&\approx
\begin{pmatrix}
0 & -e^{-i\phi}\frac{m_i}{2p}\epsilon^*_\mu \\
-e^{+i\phi}\frac{m_j}{2p}\epsilon_\mu & \bar{n}_\mu
\end{pmatrix}\,,
\end{align}
\end{subequations}
where we use the notation introduced in Eq.\,\eqref{spinflavorstruct}.
As an example, the $\nu\nu$ term, Eq.~\eqref{eq:contraction-nunu},
reads explicitly
\begin{subequations}
\label{gnuPLexpl}
\begin{align}
\framebox{\hbox to 0pt{$\vphantom{+}$}$--$}_{~ij}^{~\nu\nu}&=
\begin{pmatrix}
1 & 1 & 1\\
1 & 1 & 1\\
1 & 1 & 1
\end{pmatrix}
\,n_\mu\,,\\
\framebox{$-+$}_{~ij}^{~\nu\nu}&=
- \frac{1}{2p}
\begin{pmatrix}
m_1 & m_2 & m_3\\
m_1 & m_2 & m_3\\
m_1 & m_2 & m_3\\
\end{pmatrix}
e^{+i\phi}\epsilon^*_\mu\,,\\
\framebox{$+-$}_{~ij}^{~\nu\nu}&=
-\frac{1}{2p}
\begin{pmatrix}
m_1 & m_1 & m_1\\
m_2 & m_2 & m_2\\
m_3 & m_3 & m_3\\
\end{pmatrix}
e^{-i\phi}\epsilon_\mu\,,\\
\framebox{$++$}_{~ij}^{~\nu\nu}&=0
\,.
\end{align}
\end{subequations}
These results agree with those
obtained in Ref.~\cite{Serreau:2014cfa}.

\section{\label{sec:majorana}Majorana neutrino}

From a theoretical perspective, it is quite natural for neutrino
masses to be of Majorana type.  In this case, the two helicity states
of a given family coincide with the $\nu$ and $\bar\nu$ states, the
mass term violates lepton number, and there are no sterile degrees of
freedom. We work out the modifications of the results of the previous
section for the Majorana case, concentrating again on technical
issues.

\subsection{Two-point correlators and kinetic equations}

In the Majorana case, the momentum decomposition of the neutrino field
looks the same as for the Dirac case
Eq.~\eqref{eq:Dirac-decomposition}. However, because there are no
independent antiparticle degrees of freedom, the field mode $\vec p$
has the simpler form
\begin{align}
\label{nudecompositionmaj}
\nu_{i,s}(t,\vec{p})&= a_{i,s}(t,\vec{p})u_{i,s}(\vec{p})
+a^\dagger_{i,s}(t,-\vec{p})v_{i,s}(-\vec{p})\,.
\end{align}
The creation and annihilation operators satisfy the same anticommutation relations
Eq.~\eqref{AnticommRels} and the bispinors are the same as in the
Dirac case.

The definitions of the two-point correlation functions are different
because of the different particle content,
\begin{subequations}
\label{MajCorrelators}
\begin{align}
(2\pi)^3\delta(\vec{p}{-}\vec{k})&\rho_{ij,sh} (\vec{p})
=\langle a^\dagger_{j,h}(+\vec{k}) a_{i,s}(+\vec{p})\rangle\,,\\
(2\pi)^3\delta(\vec{p}{-}\vec{k})&\bar{\rho}_{ij,sh} (\vec{p})
=\langle a^\dagger_{i,s}(-\vec{p}) a_{j,h}(-\vec{k})\rangle\,,\\
(2\pi)^3\delta(\vec{p}{-}\vec{k})&\kappa_{ij,sh} (\vec{p})
=\langle a_{j,h}(-\vec{k}) a_{i,s}(+\vec{p})\rangle\,,\\
(2\pi)^3\delta(\vec{p}{-}\vec{k})&\kappa^\dagger_{ij,sh} (\vec{p})
=\langle a^\dagger_{j,h}(+\vec{p}) a^\dagger_{i,s}(-\vec{k})\rangle\,,
\end{align}
\end{subequations}
where all operators are taken at the same time $t$. In the Dirac case,
$\kappa^\dagger$ has no additional information relative to
$\kappa$. Here we have additional redundancies
\begin{subequations}
\label{RhoMajProperties}
\begin{align}
\bar{\rho}_{ij,sh} (t,\vec{p})&=\rho_{ji,hs} (t,-\vec{p})\,,\\
\kappa_{ij,sh} (t,\vec{p})&=-\kappa_{ji,hs} (t,-\vec{p})\,,
\end{align}
\end{subequations}
which reflect that Majorana neutrinos have half as many  degrees of freedom as
Dirac ones. Note that in the Majorana case, the pair correlations violate
total lepton number.

The mean-field Hamiltonian, bilinear in the neutrino
creation and annihilation operators, has the same form Eq.~\eqref{Heff} as in
the Dirac case. However, as we will demonstrate below,  the kernel has a more general structure,
\begin{align}
\label{MajKernel}
\Gamma_{ij} = \gamma_\mu P_L V^\mu_{ij} + P_L V^R_{ij}+ P_R V^L_{ij}\,.
\end{align}
The first piece, $V^\mu_{ij}$, is defined as in Eq.~\eqref{SigmaDirac}.
In addition, there are two scalar pieces
\begin{align}
\label{Vpot}
V^{L,R}_{ij}=\sqrt{2}\GF I^{L,R}_{ij}\,,
\end{align}
depending, as we will see, on the left-chiral and right-chiral
neutrino mean-field scalar background
\begin{align}
\label{VLVRdef}
I^{L,R}_{ij} = \langle \bar{\nu}_j P_{L,R} \nu_i\rangle\,.
\end{align}
These scalar pieces are missing in the previous literature.\footnote{In a private 
communication, the authors of Ref.~\cite{Serreau:2014cfa} agree that these terms 
should indeed be present in the Majorana case. Of course, the presence of these 
terms does not modify the overall structure of the kinetic equations.}  Their
explicit form in terms of the density matrices and pair correlators will
be given in Eq.~\eqref{IL}.

The mean-field Hamiltonian can be written in a form similar to Eq.~\eqref{HeffCrAn},
\begin{align}
\label{HeffCrAnMaj}
H_{\rm mf}=  \int_{\vec{p},s,h}& \Bigl[a^\dagger_{i,s}(\vec{p}) \Gamma^{\nu\nu}_{ij,sh}(\vec{p}) a_{j,h}(\vec{p})\nonumber\\[-4pt]
&{}+a^\dagger_{i,s}(\vec{p}) \Gamma^{\nu\bar\nu}_{ij,sh}(\vec{p}) a^\dagger_{j,h}(-\vec{p})\nonumber\\[2pt]
&{}+a_{i,s}(-\vec{p}) \Gamma^{\bar\nu\nu}_{ij,sh}(\vec{p}) a_{j,h}(\vec{p})\nonumber\\
&{}+a_{i,s}(-\vec{p}) \Gamma^{\bar\nu\bar\nu}_{ij,sh}(\vec{p}) a^\dagger_{j,h}(-\vec{p})\Bigr]\,,
\end{align}
where the matrices $\Gamma^{\nu\nu}$, $\Gamma^{\nu\bar\nu}$, etc.\ are the spinor
contractions defined in Eq.\,\eqref{GammasDef}. Using the Heisenberg equation of
motion with the Hamiltonian Eq.~\eqref{HeffCrAnMaj} one  recovers the  equation of
motion $i\dot{\sf R}=[{\sf H},{\sf R}]$, where $\sf R$ and $\sf H$ have the same
structure as in Eq.\,\eqref{RHstructure}. The components of the effective Hamiltonian
now read \cite{Serreau:2014cfa}
\begin{subequations}
\label{GammasMajDef}
\begin{align}
{\sf H}^{\nu\nu}_{ij,sh}(\vec{p})& = \Gamma^{\nu\nu}_{ij,sh}(\vec{p}) -
\Gamma^{\bar\nu\bar\nu}_{ji,hs}(-\vec{p})\,,\\
\label{GammanubarnuMaj}
{\sf H}^{\nu\bar\nu}_{ij,sh}(\vec{p})& = \Gamma^{\nu\bar\nu}_{ij,sh}(\vec{p}) -
\Gamma^{\nu\bar\nu}_{ji,hs}(-\vec{p})\,,\\
\label{GammabarnunuMaj}
{\sf H}^{\bar\nu\nu}_{ij,sh}(\vec{p})& = \Gamma^{\bar\nu\nu}_{ij,sh}(\vec{p}) -
\Gamma^{\bar\nu\nu}_{ji,hs}(-\vec{p})\,,\\
{\sf H}^{\bar\nu\bar\nu}_{ij,sh}(\vec{p})& = \Gamma^{\bar\nu\bar\nu}_{ij,sh}(\vec{p}) -
\Gamma^{\nu\nu}_{ji,hs}(-\vec{p})\,.
\end{align}
\end{subequations}
Not all of these components are independent. In particular
\begin{subequations}
\label{GammaMajProperties}
\begin{align}
{\sf H}^{\bar\nu\bar\nu}_{ij,sh}(\vec{p})=-{\sf H}^{\nu\nu}_{ji,hs}(-\vec{p})\,,\\
{\sf H}^{\nu\bar\nu}_{ij,sh}(\vec{p})=-{\sf H}^{\nu\bar\nu}_{ji,hs}(-\vec{p})\,,
\end{align}
\end{subequations}
so only two of the four submatrices of $\sf H$ are independent.

\subsection{Neutrino-neutrino mean-field Hamiltonian}

The Majorana neutrino interaction with matter is described by the same charged-
and neutral-current Hamiltonian densities Eqs.~\eqref{HCC} and~\eqref{HNC} which lead
to the same mean-field currents of electrons and nucleons---see Eqs.~\eqref{HCCeff}
and \eqref{HNCeff}.

The neutrino-neutrino interaction in the Majorana case is also described by Eq.~\eqref{Hself}.
However, Majorana neutrinos violate lepton-number conservation,
and in addition to the four lepton-number-conserving
combinations considered in Sec.~\ref{sec:dirac} one should also take into account the
lepton-number-violating combinations \smash{$\langle\nu_i\nu_j\rangle$} and
\smash{$\langle\bar\nu_i \bar{\nu}_j\rangle$} which were not included in the
previous literature.

To calculate these additional contractions, we use the definition of the charge-conjugate
field \smash{$\nu^c\equiv C\bar{\nu}^T$}, where $C$ is the charge-conjugation matrix which
has the property  $C^T C=1$. Using this definition
\smash{$\bar{\nu}=(\nu^c)^TC$} and \smash{$\nu=C\gamma^0(\nu^c)^*$}, which further
implies  \smash{$\bar{\nu}\gamma^\mu P_L\nu=-\overline{\nu^c}\gamma^\mu P_R \nu^c$}.
Therefore, we can rewrite the Hamiltonian as \smash{$-\bigl[\bar{\nu}_i\gamma^\mu P_L\nu_i\bigr]
\bigl[\overline{\nu^c_j}\gamma_\mu P_R \nu^c_j\bigr]$}
in Eq.~\eqref{Hself}. The Fierz identity \cite{Nishi:2004st} $(\gamma^\mu P_L)[\gamma_\mu P_R]=2(P_R][P_L)$
further allows us to rewrite it as \smash{$2\bigl[\bar{\nu}_i P_R\nu^c_j\bigr]
\bigl[\overline{\nu^c_j} P_L\nu_i\bigr]$}, where another sign change was induced by
anticommuting the neutrino fields. Taking the expectation value of one of the
square brackets we obtain for the new contribution to the mean-field Hamiltonian density
\begin{align}
\label{Heff3}
{\cal H}_{\rm mf}^{\nu\nu}= \sqrt{2}\GF\sum\limits_{ij}
\bigl(&\bigl[\bar{\nu}_i P_R\nu^c_j\bigr]I^L_{ij}
+\bigl[\overline{\nu^c_i} P_L\nu_j\bigr]I^R_{ij} \bigr)\,.
\end{align}
These new terms supplement the expression for the effective
Majorana Hamiltonian obtained in the previous literature \cite{Serreau:2014cfa}.
In Appendix~\ref{sec:contractions} we reproduce this result using
two-component notation.

Two comments are in order here. First,
for Majorana fer\-mi\-ons $\nu^c=\nu$ and therefore the resulting contribution to the
kernel reduces to the last two terms in Eq.\,\eqref{MajKernel}, while the definition
of left- and right-chiral neutrino backgrounds reduces to Eq.\,\eqref{Vpot}.
Second, \smash{$\bar{\nu} P_R\nu^c=\bar{\nu}_L\nu_L^c$} and
\smash{$\overline{\nu^c} P_L\nu=\overline{\nu^c_L}\nu_L$},
where \smash{$\nu_L\equiv  P_L\nu$}, which are nothing but
components of the Majorana mass term.

\subsection{\label{EffHamMaj}Components of the Hamiltonian matrix \boldmath{$\sf H$}}

The new contributions stemming from neutrino-neutrino interactions can be expressed
in terms of the (anti)particle densities and pair correlators,
\begin{align}
\label{IL}
I^L_{ij}=\int_{\vec{p},s,h}\Bigl[&\,(P_L)^{\nu\nu}_{ji,hs}\rho_{ij,sh}
+(P_L)^{\bar\nu\bar\nu}_{ji,hs}(\delta_{ij}\delta_{sh}-\bar{\rho}_{ij,sh})\nonumber\\[-5pt]
&{}+(P_L)^{\bar\nu\nu}_{ji,hs}\kappa_{ij,hs}
+(P_L)^{\nu\bar\nu}_{ji,hs}\kappa^\dagger_{ij,sh}\Bigr]\,,
\end{align}
where we have again suppressed the common arguments $\vec p$ and $(t,\vec{p})$. The notation
for the scalar contractions 
$(P_L)^{\nu\nu}_{ij,sh}$ etc.\ is analogous to Eq.~\eqref{eq:contractiondefinition}, except
that now there is no $\gamma^\mu$ included.

To lowest order in the small neutrino masses we find, using the explicit form of the chiral spinors
of Appendix~\ref{sec:spinorproducts},
\begin{subequations}
\label{XiL}
\begin{align}
(P_L)^{\nu\nu}_{ij,sh}&\approx
\begin{pmatrix}
\frac{m_i}{2p} & 0 \\
0 &\frac{m_j}{2p}
\end{pmatrix}\,,\\
(P_L)^{\nu\bar\nu}_{ij,sh}
&\approx
\begin{pmatrix}
0 & 0 \\
0 & -e^{-i\phi}
\end{pmatrix}\,,\\
(P_L)^{\bar\nu\nu}_{ij,sh}
&\approx
\begin{pmatrix}
e^{-i\phi} & 0 \\
0 & 0
\end{pmatrix}\,,\\
(P_L)^{\bar\nu\bar\nu}_{ij,sh}&\approx
\begin{pmatrix}
-\frac{m_j}{2p} & 0 \\
0 & -\frac{m_i}{2p}
\end{pmatrix}\,.
\end{align}
\end{subequations}
The components of $(P_R)$ can be obtained from these results using the relations
$(P_R)^{\nu\nu}_{ij,sh}=[(P_L)^{\nu\nu}_{ji,hs}]^*$ and
$(P_R)^{\nu\bar\nu}_{ij,sh}=[(P_L)^{\bar\nu\nu}_{ji,hs}]^*$, as well
as similar relations for the remaining two components.

Using the definitions Eq.~\eqref{GammasMajDef} combined with
Eq.~\eqref{Xinunu} and the corresponding definition for the scalar case
we obtain for the $\nu\nu$ component of $\sf H$
\begin{align}
\label{GammaMnunu}
{\sf H}^{\nu\nu}_{ij,sh}&(\vec{p})=\delta_{sh}\delta_{ij} E_i\nonumber\\
&+(\gamma_\mu P_L)^{\nu\nu}_{ij,sh}(\vec{p})V^\mu_{ij}-
(\gamma_\mu P_L)^{\bar\nu\bar\nu}_{ji,hs}(-\vec{p})V^\mu_{ji}\nonumber\\
&+(P_L)^{\nu\nu}_{ij,sh}(\vec{p})V^R_{ij}-
(P_L)^{\bar\nu\bar\nu}_{ji,hs}(-\vec{p})V^R_{ji}\nonumber\\
&+(P_R)^{\nu\nu}_{ij,sh}(\vec{p})V^L_{ij}-
(P_R)^{\bar\nu\bar\nu}_{ji,hs}(-\vec{p})V^L_{ji}\,.
\end{align}
The second line generalizes the Dirac result of Eq.~\eqref{GammaDiracnunu}
to the Majorana case and has been obtained in Ref.~\cite{Serreau:2014cfa}. The third and
fourth lines stem from the the contractions Eq.~\eqref{Heff3} and supplement the
previous results.

The $\bar\nu\bar\nu$ term follows from the identity Eq.~\eqref{GammaMajProperties}.
For the $\nu\bar\nu$ component we find
\begin{align}
\label{GammaMnunubar}
{\sf H}^{\nu\bar\nu}_{ij,sh}(\vec{p})
&=(\gamma_\mu P_L)^{\nu\bar\nu}_{ij,sh}(\vec{p})V^\mu_{ij}-
(\gamma_\mu P_L)^{\nu\bar\nu}_{ji,hs}(-\vec{p})V^\mu_{ji}\nonumber\\
&+(P_L)^{\nu\bar\nu}_{ij,sh}(\vec{p})V^R_{ij}-
(P_L)^{\nu\bar\nu}_{ji,hs}(-\vec{p})V^R_{ji}\nonumber\\
&+(P_R)^{\nu\bar\nu}_{ij,sh}(\vec{p})V^L_{ij}-
(P_R)^{\nu\bar\nu}_{ji,hs}(-\vec{p})V^L_{ji}\,.
\end{align}
The $\bar\nu\nu$ component follows from replacing
$\nu\bar\nu$ with $\bar\nu\nu$ everywhere in this result.

An inspection of Eqs.~\eqref{IL} and \eqref{XiL} shows that the last two lines of Eq.~\eqref{GammaMnunu}
contain terms proportional to $\kappa$ and $\kappa^\dagger$ that are linear
in the neutrino masses, and additionally terms quadratic in the neutrino masses which we neglect here.

A peculiar feature of Eq.~\eqref{GammaMnunubar} is
that its last two lines contain terms proportional to $\kappa$ and $\kappa^\dagger$ that
are not suppressed by the neutrino masses and therefore do not vanish when we set
the masses to zero.  This is somewhat surprising because we expect that
Dirac and Majorana neutrinos are equivalent for $m_\nu\to0$.
Therefore, the components of ${\sf H}$ must coincide in this limit.
We return to this question in Sec.~\ref{sec:weyl},
where we study the case of massless two-component neutrinos and demonstrate that
in the massless limit  these additional terms, which are proportional to the 
lepton-number-violating correlators, are not produced if they are zero initially.

On the other hand, one important finding of our paper
is that for a Majorana neutrino with an arbitrary small mass, lepton-number-violating
correlators are automatically produced and, in turn, induce the additional scalar background terms of
the mean-field Hamiltonian which then affect  the dynamics of the density matrices.

\section{\label{sec:weyl}Weyl neutrino}
In the previous section we have found that the additional scalar contributions
to the mean-field Hamiltonian, that naturally arise for Majorana neutrinos,
do not vanish   in the  massless limit. This is somewhat surprising
because we expect no difference between Dirac and Majorana neutrinos in this
case. To clarify this paradox we study a single generation of
massless neutrinos. The equations presented in this section will also be
used later to study particle-antiparticle coherence.

\subsection{Standard two-point correlators and kinetic equations}

In the Weyl case, the momentum decomposition of the neutrino field
looks the same  as for the Dirac case Eq.~\eqref{eq:Dirac-decomposition}.
However, because a Weyl fermion has only two degrees of freedom the
field mode $\vec p$ does not carry a spin index,
\begin{align}
\label{nudecompositionweil}
\nu(t,\vec{p})&= a(t,\vec{p})u_{_-}(\vec{p})+ b^\dagger(t,-\vec{p})v_{_+}(-\vec{p})\,.
\end{align}
It is automatically left-chiral because the right-chiral components of the chiral
spinors $u_{_-}(\vec{p})$ and $v_{_+}(-\vec{p})$ vanish in the massless limit,
see Appendix \ref{sec:spinorproducts}.

If we require lepton-number conservation then the only correlators that we can
define are
\begin{subequations}
\label{RhoKappaWeylCons}
\begin{align}
(2\pi)^3\delta(\vec{p}-\vec{k})\rho_{{\smi\smi}} (\vec{p})&=
\langle a^\dagger(\vec{k}) a(\vec{p})\rangle\,,\\
(2\pi)^3\delta(\vec{p}-\vec{k})\bar{\rho}_{{\spi\spi}} (\vec{p})&=
\langle b^\dagger(-\vec{k}) b(-\vec{p})\rangle\,,\\
(2\pi)^3\delta(\vec{p}-\vec{k})\kappa_{{\smi\spi}} (\vec{p})&=
\langle b(-\vec{k}) a(\vec{p})\rangle\,,\\
(2\pi)^3\delta(\vec{p}-\vec{k})\kappa^\dagger_{{\spi\smi}} (\vec{p})&=
\langle a^\dagger(\vec{p}) b^\dagger(-\vec{k})\rangle\,.
\end{align}
\end{subequations}
Note that we keep   helicity indices in   these definitions to distinguish
the lepton-number-conserving correlators from the lepton-number-violating ones,
which we introduce below.

We can extract the explicit form of the kinetic equations for these correlators
from Eq.\,\eqref{eq:EOM1},
\begin{subequations}
\label{RhoKappaEqWeyl}
\begin{align}
\label{RhoKappaEqWeylnunu}
i\dot{\rho}_{{\smi\smi}}&={\sf H}^{\nu\nu}_{{\smi\smi}}\rho_{{\smi\smi}}-
\rho_{{\smi\smi}}{\sf H}^{\nu\nu}_{{\smi\smi}}+{\sf H}^{\nu\bar\nu}_{{\smi\spi}}\kappa^\dagger_{{\spi\smi}}-
\kappa_{{\smi\spi}}{\sf H}^{\bar\nu\nu}_{{\spi\smi}}\,,\\
\label{RhoKappaEqWeylnubarnubar}
i\dot{\bar\rho}_{{\spi\spi}}&={\sf H}^{\bar\nu\bar\nu}_{{\spi\spi}}\bar{\rho}_{{\spi\spi}}-
\bar{\rho}_{{\spi\spi}}{\sf H}^{\bar\nu\bar\nu}_{{\spi\spi}}
-{\sf H}^{\bar\nu\nu}_{{\spi\smi}}\kappa_{{\smi\spi}}
+\kappa^\dagger_{{\spi\smi}}{\sf H}^{\nu\bar\nu}_{{\smi\spi}}\,,\\
\label{KappaEq}
i\dot{\kappa}_{{\smi\spi}}&={\sf H}^{\nu\nu}_{{\smi\smi}}\kappa_{{\smi\spi}}-
\kappa_{{\smi\spi}}{\sf H}^{\bar\nu\bar\nu}_{{\spi\spi}}
-{\sf H}^{\nu\bar\nu}_{{\smi\spi}}\bar{\rho}_{{\spi\spi}}-
\rho_{{\smi\smi}}{\sf H}^{\nu\bar\nu}_{{\smi\spi}}\nonumber\\
&+{\sf H}^{\nu\bar\nu}_{{\smi\spi}}\,,
\end{align}
\end{subequations}
where  we  omit the arguments $(t,\vec{p})$, which are common to all the functions,
to shorten the notation. Note that for a single neutrino generation the first two 
terms in Eqs.~\eqref{RhoKappaEqWeylnunu} and \eqref{RhoKappaEqWeylnubarnubar} cancel 
each other and we have retained them only to keep the resemblance with the general 
form of the kinetic equations.

A peculiar feature of Eq.~\eqref{KappaEq} is that $\kappa$,
i.e.\ the coherence between $|00\rangle$ and $|11\rangle$ states, is automatically induced
provided that the mean-field Hamiltonian $\sf H$ has nonzero off-diagonals. The
off-diagonals can be induced even if all neutrino two-point functions are zero initially
by, for instance, a transverse neutron current.

The explicit form of the mean-field Hamiltonian can be obtained from
Eq.~\eqref{Xinunu} by setting the masses to zero,
\begin{subequations}
\label{GammaWeyl}
\begin{align}
\label{GammaWeylnunu}
{\sf H}^{\nu\nu}_{{\smi\smi}}(\vec{p})&=E+V^0-\hat{\vec p}\vec{V}\,,\\
\label{GammaNuBarnu}
{\sf H}^{\nu\bar\nu}_{{\smi\spi}}(\vec{p})&=-\hat{\vec\epsilon}^* \vec{V}\,,\\
{\sf H}^{\bar\nu\nu}_{{\spi\smi}}(\vec{p})&=-\hat{\vec\epsilon}\, \vec{V}\,,\\
{\sf H}^{\bar\nu\bar\nu}_{{\spi\spi}}(\vec{p})&=- E+V^0+\hat{\vec p}\vec{V}\,,
\end{align}
\end{subequations}
where $E=\abs{\vec{p}}$. Note that the $\hat{\vec p}\vec{V}$ term in Eq.~\eqref{GammaWeylnunu}  accounts for
the enhancement (suppression) of the mean-field potential for the  matter flowing
antiparallel (parallel) to the neutrino momentum. This has been  pointed out in
Ref.~\cite{Grigoriev:2002zr}.

It remains to express the neutrino current  $I^\mu$ in terms of the density
matrices and pair correlations. For its time component we obtain from
Eq.\,\eqref{InuExplicit}, \smash{$I^0=\int_{\vec p} \ell$},
where $\ell(t,\vec{p})\equiv \rho(t,\vec{p})-\bar{\rho}(t,-\vec{p})$
has the meaning of lepton number in mode $\vec p$.  For the spatial
components we find
\begin{align}
\label{Inuvec}
\vec{I}&=\int_{\vec p}\bigl[
\hat{\vec p}\,\ell+\hat{\vec \epsilon}\,\kappa
+\hat{\vec \epsilon}^*\kappa^\dagger\bigr]\,,
\end{align}
which coincides with the result of Ref.~\cite{Serreau:2014cfa}.

\subsection{Lepton-number-violating correlators and kinetic equations}

If we allow for \smash{$\langle\nu\nu\rangle$} and
\smash{$\langle\bar\nu \bar{\nu}\rangle$} contractions then, similarly to the
Majorana case, the mean-field Hamiltonian receives contributions of the type
Eq.\,\eqref{Heff3}.  Because Weyl fields satisfy the condition $P_L\nu=\nu$
we can rewrite Eq.\,\eqref{Heff3} as
\begin{align}
\label{HeffWeyl}
\mathcal{H}^{\nu\nu}_{\rm mf}= \sqrt{2}\GF\!\sum\limits
\bigl(\bigl[\bar{\nu} \nu^c\bigr]I^L
\!+\!\bigl[\overline{\nu^c} \nu\bigr]I^R \bigr)\,
\end{align}
(see Sec.\,\ref{sec:majorana} and Appendix \ref{sec:contractions} for more details) where now
\begin{align}
\label{SigmaWeyl}
I^L= \langle \overline{\nu^c} \nu\rangle\,\quad {\rm and}\quad
I^R = \langle \bar{\nu} \nu^c\rangle\,.
\end{align}
As has been mentioned above  $\overline{\nu^c} \nu$ and $\bar{\nu} \nu^c$
have the structure of the Majorana mass term, which is known to violate lepton number.
Therefore, we expect that also for the
Weyl neutrino the mean-field Hamiltonian Eq.~\eqref{HeffWeyl} leads to lepton number violation. However, for
Weyl neutrinos the inclusion of these additional terms is somewhat artificial because,
as we show below,  these correlations  are not produced if they are zero initially.
They are considered here to better understand the Majorana case, where they are
naturally produced by the lepton-number-violating interactions.

The contribution of Eq.\,\eqref{HeffWeyl} to the mean-field Hamiltonian is given by
\begin{align}
\label{HLeptViol}
H_{\rm mf}= \int_{\vec{p}} \bigl[&a^\dagger(\vec{p}) \Gamma^{\nu\bar\nu}_{{\smi\smi}}(\vec{p}) a^\dagger(-\vec{p})\nonumber\\[-2mm]
+\, &b^\dagger(\vec{p}) \Gamma^{\nu\bar\nu}_{{\spi\spi}}(\vec{p}) b^\dagger(-\vec{p})\nonumber\\
+\, &a(-\vec{p}) \Gamma^{\bar\nu\nu}_{{\smi\smi}}(\vec{p}) a(\vec{p})\nonumber\\
+\, &b(-\vec{p}) \Gamma^{\bar\nu\nu}_{{\spi\spi}}(\vec{p}) b(\vec{p})\bigr]\,,
\end{align}
and strongly resembles the mean-field Hamiltonian of Majorana neutrinos
Eq.\,\eqref{HeffCrAnMaj}.
From the structure of Eq.\,\eqref{HLeptViol} it is evident  that, as expected,
it leads to the violation of lepton number. To take this into account we
are forced to introduce the following lepton-number-violating correlators,
\begin{subequations}
\label{RhoKappaWeylViol}
\begin{align}
(2\pi)^3\delta(\vec{p}-\vec{k})\kappa_{{\smi\smi}} (\vec{p})&=
\langle a(-\vec{k}) a(\vec{p})\rangle\,,\\
(2\pi)^3\delta(\vec{p}-\vec{k})\kappa_{{\spi\spi}} (\vec{p})&=
\langle b(-\vec{k}) b(\vec{p})\rangle\,,\\
(2\pi)^3\delta(\vec{p}-\vec{k})\kappa^\dagger_{{\smi\smi}} (\vec{p})&=
\langle a^\dagger(\vec{p}) a^\dagger(-\vec{k})\rangle\,,\\
(2\pi)^3\delta(\vec{p}-\vec{k})\kappa^\dagger_{{\spi\spi}} (\vec{p})&=
\langle b^\dagger(\vec{p}) b^\dagger(-\vec{k})\rangle\,,
\end{align}
\end{subequations}
which also resemble  the Majorana definitions  Eq.\,\eqref{MajCorrelators}.
These correlators are dictated by the structure of
the Hamiltonian Eq.\,\eqref{HLeptViol} and are the only lepton-number-violating
correlators we consider in this section. If we wanted to consider all other
possible correlators we would be back to the Majorana case with zero neutrino
masses.

The lepton-number-violating correlators contribute to the dynamics of the
lepton-number-conserving ones,
\begin{subequations}
\label{RhoKappaEqLeptViol}
\begin{align}
i\dot{\rho}_{{\smi\smi}}&=\ldots\nonumber\\
&+\bigl(\Gamma^{\nu\bar\nu}_{{\smi\smi}}-[\Gamma^{\nu\bar\nu}_{{\smi\smi}}]^T\bigr)\kappa^\dagger_{{\smi\smi}}
-\kappa_{{\smi\smi}}\bigl(\Gamma^{\bar\nu\nu}_{{\smi\smi}}-[\Gamma^{\bar\nu\nu}_{{\smi\smi}}]^T\bigr)\,,\\
i\dot{\bar\rho}_{{\spi\spi}}&=\ldots\nonumber\\
&-\bigl(\Gamma^{\bar\nu\nu}_{{\spi\spi}}-[\Gamma^{\bar\nu\nu}_{{\spi\spi}}]^T\bigr)\kappa_{{\spi\spi}}
+\kappa^\dagger_{{\spi\spi}}\bigl(\Gamma^{\nu\bar\nu}_{{\spi\spi}}-[\Gamma^{\nu\bar\nu}_{{\spi\spi}}]^T\bigr)\,,
\end{align}
\end{subequations}
where ellipses denote   terms on the right-hand side of Eq.\,\eqref{RhoKappaEqWeyl},
and the superscript $T$ stands for transposition of the flavor and helicity indices, as well as
inversion of the momentum. Comparing Eq.\,\eqref{RhoKappaEqLeptViol} with
Eqs.\,\eqref{GammanubarnuMaj} and \eqref{GammabarnunuMaj} we see that we automatically
recover the ``Majorana'' definitions of the Hamiltonian matrix.
Note that to avoid confusion with the definitions of the elements of the mean-field
Hamiltonian, which are different for Dirac and Majorana neutrinos, we write the
right-hand side of Eq.\,\eqref{RhoKappaEqLeptViol} directly in terms of spinor contractions
defined in Eq.\,\eqref{GammasDef}.
The dynamics of $\kappa_{{\smi\spi}}$, see Eq.\,\eqref{KappaEq}, does not receive
any corrections. The reason is that the components of the mean-field Hamiltonian
needed to form the right spin combination with the lepton-number-violating correlators
in Eq.~\eqref{KappaEq}  are zero for Weyl neutrinos. The kinetic equations for the
lepton-number-violating pair correlations read
\begin{subequations}
\begin{align}
i\dot{\kappa}_{{\smi\smi}}&=
\Gamma^{\nu\nu}_{{\smi\smi}}\kappa_{{\smi\smi}}-
\kappa_{{\smi\smi}}\bigl(-[\Gamma^{\nu\nu}_{{\smi\smi}}]^T\bigr)
-\rho_{{\smi\smi}}\bigl(\Gamma^{\nu\bar\nu}_{{\smi\smi}}-[\Gamma^{\nu\bar\nu}_{{\smi\smi}}]^T\bigr)\nonumber\\
&-\bigl(\Gamma^{\nu\bar\nu}_{{\smi\smi}}-[\Gamma^{\nu\bar\nu}_{{\smi\smi}}]^T\bigr)[\rho_{{\smi\smi}}]^T
+\bigl(\Gamma^{\nu\bar\nu}_{{\smi\smi}}-[\Gamma^{\nu\bar\nu}_{{\smi\smi}}]^T\bigr)\,,\\
i\dot{\kappa}_{{\spi\spi}}&= \bigl(-[\Gamma^{\bar\nu\bar\nu}_{{\spi\spi}}]^T\bigr)\kappa_{{\spi\spi}}
-\kappa_{{\spi\spi}}\Gamma^{\bar\nu\bar\nu}_{{\spi\spi}}
-\bigl(\Gamma^{\nu\bar\nu}_{{\spi\spi}}-[\Gamma^{\nu\bar\nu}_{{\spi\spi}}]^T\bigr)\bar{\rho}_{{\spi\spi}}\nonumber\\
&-[\bar{\rho}_{{\spi\spi}}]^T\bigl(\Gamma^{\nu\bar\nu}_{{\spi\spi}}-[\Gamma^{\nu\bar\nu}_{{\spi\spi}}]^T\bigr)
+\bigl(\Gamma^{\nu\bar\nu}_{{\spi\spi}}-[\Gamma^{\nu\bar\nu}_{{\spi\spi}}]^T\bigr)\,.
\end{align}
\end{subequations}
Their form can be guessed from  Eq.\,\eqref{KappaEq} by replacing components
of the mean-field Hamiltonian with their ``Majorana'' counterparts, taking into account
that \smash{$\Gamma^{\bar\nu\bar\nu}_{{\smi\smi}}=\Gamma^{\nu\nu}_{{\spi\spi}}=0$}, and
replacing $\bar{\rho}_{{\smi\smi}}$ by \smash{$[\rho_{{\smi\smi}}]^T$} as well as
$\rho_{{\spi\spi}}$ by \smash{$[\bar{\rho}_{{\spi\spi}}]^T$}.

Using the explicit form of the chiral spinors (see Appendix \ref{sec:spinorproducts})
we obtain
\begin{subequations}
\label{GammaWeylnunuCorr}
\begin{align}
\Gamma^{\nu\bar\nu}_{{\smi\smi}}(\vec{p})&=+e^{+i\phi}V^L\,,\\
\Gamma^{\nu\bar\nu}_{{\spi\spi}}(\vec{p})&=-e^{-i\phi}V^R\,,\\
\Gamma^{\bar\nu\nu}_{{\smi\smi}}(\vec{p})&=+e^{-i\phi}V^R\,,\\
\Gamma^{\bar\nu\nu}_{{\spi\spi}}(\vec{p})&=-e^{+i\phi}V^L\,,
\end{align}
\end{subequations}
where \smash{$V^{L(R)}=\sqrt{2}\GF I^{L(R)}$} are defined analogously to Eq.\,\eqref{Vpot}.
Let us now recall that \smash{$I^L$} and \smash{$I^R$} are produced
only by neutrino self-in\-teractions and are proportional to the lepton-number-violating pair correlations,
\begin{align}
I^L=\int_{\vec p} e^{-i\phi}\bigl[\kappa_{{\smi\smi}}-\kappa^\dagger_{{\spi\spi}}\bigr]\,,
\end{align}
and a similar expression for \smash{$I^{R}$}. Thus, if the lepton-number-violating 
correlators are zero initially,
then the components in Eq.~\eqref{GammaWeylnunuCorr} are zero and $\kappa_{\smi\smi}$ and
$\kappa_{\spi\spi}$ remain zero in the course of the system's evolution. For this reason
for Weyl neutrinos the inclusion of lepton-number-violating correlators
is rather artificial because they could only exist if they were put in by hand initially.

This observation
explains why similar contributions do not vanish for Majorana neutrinos in the limit of
zero neutrino masses. While such lepton-number-violating correlators can be introduced
by hand as an initial condition, they can  dynamically evolve only in the presence
of a nonvanishing Majorana mass.

\section{\label{sec:magnetic}Electromagnetic background fields}

A supernova environment is characterized not only by matter currents, but also by strong magnetic fields. Electromagnetic fields polarize both background media and the vacuum. Although neutrinos do not couple directly to the electromagnetic fields, they feel the induced polarization. The coupling to a polarized background medium has been treated in the previous sections. We now turn to the interaction with the vacuum polarization.

The effect of vacuum polarization is described by electromagnetic form factors. The most prominent examples, the magnetic and electric dipole moments, are inevitable for massive neutrinos and have to be included to obtain consistent evolution equations linear in the neutrino mass. The main effects of electromagnetic fields are spin and spin-flavor oscillations, which can be significant. We treat Dirac and Majorana neutrinos separately.

\subsection{General vertex structure}

The coupling of neutrinos to an external vector potential $A^\mu$ can be written as an effective vertex
$\mathcal{H}^{\rm em}= A_{\mu}\bar \nu \Gamma^{\mu} \nu$, where $\Gamma^{\mu}$ contains all
irreducible combinations of Lorentz vectors and pseudovectors generated by external momenta and
Dirac matrices. Neglecting a hypothetical minicharge, in coordinate space the most general Hamiltonian
density can be reduced to
\begin{align}
\label{eq:moment}
\mathcal{H}^{\rm em}&=\frac{1}{2}F_{\mu\nu}\,\bar \nu_i \bigl( f^{ij}_M \sigma^{\mu\nu} +if^{ij}_E\sigma^{\mu\nu}\gamma_5 \bigr)\nu_j\nonumber\\
&+\partial^\nu\! F_{\mu\nu}  \, \bar \nu_i  \bigl( f^{ij}_Q\gamma^\mu +f^{ij}_A\gamma^\mu\gamma_5\bigr) \nu_j\, ,
\end{align}
where the electromagnetic field-strength tensor is defined as usual, $F^{\mu\nu}=\partial^\mu A^\nu - \partial^\nu A^\mu$, and $\sigma^{\mu\nu}=\frac{i}{2} [\gamma^{\mu},\gamma^{\nu}]$. The form factors are $f_M$ (magnetic), $f_E$ (electric), $f_Q$ (reduced charge~\cite{Giunti:2014ixa}), and $f_A$ (anapole). The form factors carry generation indices. Diagonal elements describe the usual electromagnetic properties of a neutrino in the mass basis, and reduce to electromagnetic \emph{moments} in the static limit. The off-diagonal elements describe transitions between neutrinos of different masses. Some components of the Hamiltonian matrix have been calculated in Refs.~\cite{Giunti:2014ixa, Dvornikov:2011dv}.

Maxwell's equations tell us that $\partial_\nu F^{\mu\nu}=-J^\mu_{\text{em}}$, where $J^\mu_{\text{em}}$ is some charged matter background that sources electromagnetic fields. In supernovae, the sources are electrons and protons. In the Standard Model with massless neutrinos, the value for the anapole moment has to be $f_A=-f_Q$ to reproduce the left-chiral form of the interaction. For models with neutrino masses, the Hamiltonian matrix might obtain contributions that are not purely left-chiral, but we assume that these are always small so that we can neglect them. The charge and anapole form factors then only yield radiative corrections to the left-chiral tree-level coupling in Eq.~\eqref{HCC}. We neglect these moments because we are not interested in corrections to leading-order effects. However, for completeness, we give the spinor contractions for right-chiral currents in Appendix~\ref{sec:anapole}.

\subsection{Dipole moments of Dirac neutrinos}

To study the dipole moments, we first turn to the somewhat simpler case of Dirac neutrinos. A Dirac neutrino has diagonal magnetic and electric moments. Because we assume neutrinos to carry no charge, $\mu=f_M(0)$ is defined as the magnetic moment and $\epsilon=f_E(0)$ as the electric dipole moment~\cite{Giunti:2014ixa}. In the minimal extension of the Standard Model, the magnetic moments are found to be~\cite{Shrock:1982sc}
\begin{subequations}
\begin{align}
\label{eq:mmon1}
  \mu_{ij} & = \frac{3e\sqrt{2}\GF (m_i+m_j)}{ 2(4\pi)^2}\left(\delta_{ij}-\frac{m_\tau^2}{2m_W^2}\mathcal{F}_{ij}\right)\, ,\\
  \label{eq:mmon2}\epsilon_{ij} & = i\frac{3e\sqrt{2}\GF }{2 (4\pi)^2}(m_i-m_j)\left(\frac{m_\tau^2}{2m_W^2}\right)\mathcal{F}_{ij}\, ,
\end{align}
\begin{align}
  \mathcal{F}_{ij} & =\sum_{\alpha=e,\mu,\tau} U^\dagger_{i \alpha}  \left(\frac{m_\alpha}{m_\tau}\right)^2U_{\alpha j} \, \label{fij},
\end{align}
\end{subequations}
where $m_\tau$ is the tau mass. Note that the electric dipole moment does not have a diagonal component because it would violate \textit{CP}~\cite{Giunti:2014ixa}, and that the transition electric dipole moment carries a phase relative to the transition magnetic dipole moment. Numerically, the above expressions yield for the diagonal magnetic moments
\begin{equation}\label{eq:muD}
 \mu_{ii}\simeq 3.2\times 10^{-19} \left(\frac{m_i}{{\rm{eV}}}\right)\mu_\text{B}\, ,
\end{equation}
where $\mu_\text{B}$ is the Bohr magneton. The transition moments are
\begin{subequations}
\begin{align}
 \mu_{ij}\simeq -3.9\times 10^{-23}\mathcal{F}_{ij} \left(\frac{m_i+m_j}{{\rm{eV}}}\right)\mu_\text{B}\,,\\
 \epsilon_{ij}\simeq  3.9\, i\times 10^{-23}\mathcal{F}_{ij}\left(\frac{m_i-m_j}{{\rm{eV}}}\right)\mu_\text{B}\, .
\end{align}
\end{subequations}
Note that the transition moments are much smaller than the diagonal moments due to Glashow-Iliopoulos-Maiani suppression.

\subsection{Hamiltonian matrix for Dirac neutrinos}

We treat electromagnetic effects on the same footing as background matter. To this end, we have to evaluate the components of the Hamiltonian matrix, which, for Dirac neutrinos, are equal to the spinor contractions in Eq.~\eqref{GammasDef}. For the contractions, we need to evaluate the Lorentz structure of the vertex in Eq.~\eqref{eq:moment}.

Considering only magnetic and electric form factors, the Hamiltonian  reduces to
\smash{$\frac{1}{2}F_{\mu\nu}\bar \nu_i \bigl( f^{ij}_M \sigma^{\mu\nu} +if^{ij}_E\sigma^{\mu\nu}\gamma_5 \bigr)\nu_j$,}
which depends on the electric and magnetic fields, $\vec{E}$ and $\vec{B}$, through $F^{\mu\nu}$. The Lorentz structure can be decomposed into the contractions \smash{$\left(i\gamma^0\vec{\gamma}\right)_{ij,sh}$ and $\left(\gamma^0\vec{\gamma}\gamma_5\right)_{ij,sh}$}, the latter appearing through the identity \smash{$\epsilon^{abc}\gamma^0\gamma^c\gamma_5=\sigma^{ab}$} with spatial indices $a,b,c=1,2 $ or $3$, and the asymmetric tensor \smash{$\epsilon^{abc}$}. These contractions are three-vectors that are contracted with the electric and magnetic fields. We calculate the contractions in momentum space.

Explicitly, the coupling of the magnetic field through the magnetic form factor (superscript $\mu \rm B$) has the structures
\begin{subequations}\label{eq:nunumuB}
\begin{align}
  \mathsf{H}^{\mu\text{B} \nu \nu}_{ij,sh}=&-\bigl(\gamma^0\vec{\gamma}\gamma_5\bigr)^{\nu\nu}_{ij,sh}f^{ij}_{M}(q^2)\vec{B}\, ,\\
  \mathsf{H}^{\mu\text{B} \nu \bar\nu}_{ij,sh}=&-\bigl(\gamma^0\vec{\gamma}\gamma_5\bigr)^{\nu\bar\nu}_{ij,sh}f^{ij}_{M}(l^2)\vec{B}\, ,\\
  \mathsf{H}^{\mu\text{B} \bar\nu \nu}_{ij,sh}=&-\bigl(\gamma^0\vec{\gamma}\gamma_5\bigr)^{\bar\nu\nu}_{ij,sh}f^{ij}_{M}(l^2)\vec{B}\, ,\\
  \mathsf{H}^{\mu\text{B} \bar\nu \bar\nu}_{ij,sh}=&-\bigl(\gamma^0\vec{\gamma}\gamma_5\bigr)^{\bar\nu\bar\nu}_{ij,sh}f^{ij}_{M}(q^2)\vec{B}\, ,
\end{align}
\end{subequations}
where we identify ${\sf H}^{\nu\nu}=\Gamma^{\nu\nu}$, ${\sf H}^{\nu\bar\nu}=\Gamma^{\nu\bar\nu}$ , etc.,
and  the minus sign in the metric  $g^{\mu\nu}=\text{diag}(1,-1,-1,-1)$
has already been taken care of. In Eq.~\eqref{eq:nunumuB}, the form factors still depend on the momentum transfer. For the $\nu\nu$ and $\bar\nu\bar\nu$ components, the form factors contain  $q^\mu=p^\mu_{\text{out}}-p^\mu_{\text{in}}$, where $q^\mu\to 0$ in the forward-scattering limit. These components are then proportional to the dipole moments. For the neutrino-antineutrino components of the $\mathsf{H}$ matrices, the argument of the form factor contains $l^2$ with $l^\mu =p^\mu_{\text{out}}+p^\mu_{\text{in}}$, the sum of neutrino and antineutrino momenta. In the forward-scattering limit this reduces to $l^2=(2E)^2$, and the dependence of the form factors on the four-momentum is important.

The coupling of the magnetic field to the electric form factor (superscript $\epsilon \rm B$) is
\begin{subequations}
\begin{align}
  \mathsf{H}^{\epsilon\text{B} \nu \nu}_{ij,sh}=&-\bigl(i\gamma^0\vec{\gamma}\bigr)^{\nu\nu}_{ij,sh}f^{ij}_{E}(q^2)\vec{B}\,,\\
  \mathsf{H}^{\epsilon\text{B} \nu \bar\nu}_{ij,sh}=&-\bigl(i\gamma^0\vec{\gamma}\bigr)^{\nu\bar\nu}_{ij,sh}f^{ij}_{E}(l^2)\vec{B}\,,\\
  \mathsf{H}^{\epsilon\text{B} \bar\nu \nu}_{ij,sh}=&-\bigl(i\gamma^0\vec{\gamma}\bigr)^{\bar\nu\nu}_{ij,sh}f^{ij}_{E}(l^2)\vec{B}\,,\\
  \mathsf{H}^{\epsilon\text{B} \bar\nu \bar\nu}_{ij,sh}=&-\bigl(i\gamma^0\vec{\gamma}\bigr)^{\bar\nu\bar\nu}_{ij,sh}f^{ij}_{E}(q^2)\vec{B}\,.
\end{align}
\end{subequations}
The coupling of an electric field to the magnetic form factor is
\begin{subequations}
\begin{align}
  \mathsf{H}^{\mu\text{E} \nu \nu}_{ij,sh}=&\bigl(i\gamma^0\vec{\gamma}\bigr)^{\nu\nu}_{ij,sh}f^{ij}_{M}(q^2)\vec{E}\,,\\
  \mathsf{H}^{\mu\text{E} \nu \bar\nu}_{ij,sh}=&\bigl(i\gamma^0\vec{\gamma}\bigr)^{\nu\bar\nu}_{ij,sh}f^{ij}_{M}(l^2)\vec{E}\, ,\\
  \mathsf{H}^{\mu\text{E} \bar\nu \nu}_{ij,sh}=&\bigl(i\gamma^0\vec{\gamma}\bigr)^{\bar\nu\nu}_{ij,sh}f^{ij}_{M}(l^2)\vec{E}\, ,\\
  \mathsf{H}^{\mu\text{E} \bar\nu \bar\nu}_{ij,sh}=&\bigl(i\gamma^0\vec{\gamma}\bigr)^{\bar\nu\bar\nu}_{ij,sh}f^{ij}_{M}(q^2)\vec{E}\, ,
\end{align}
\end{subequations}
which is indicated by $\mu \rm E$, and to the electric form factor, $\epsilon \rm E$,
\begin{subequations}\label{eq:nunuepE}
\begin{align}
  \mathsf{H}^{\epsilon\text{E} \nu \nu}_{ij,sh}=&-\bigl(\gamma^0\vec{\gamma}\gamma_5\bigr)^{\nu\nu}_{ij,sh}f^{ij}_{E}(q^2)\vec{E},\\
  \mathsf{H}^{\epsilon\text{E} \nu \bar\nu}_{ij,sh}=&-\bigl(\gamma^0\vec{\gamma}\gamma_5\bigr)^{\nu\bar\nu}_{ij,sh}f^{ij}_{E}(l^2)\vec{E},\\
  \mathsf{H}^{\epsilon\text{E} \bar\nu \nu}_{ij,sh}=&-\bigl(\gamma^0\vec{\gamma}\gamma_5\bigr)^{\bar\nu\nu}_{ij,sh}f^{ij}_{E}(l^2)\vec{E},\\
  \mathsf{H}^{\epsilon\text{E} \bar\nu \bar\nu}_{ij,sh}=&-\bigl(\gamma^0\vec{\gamma}\gamma_5\bigr)^{\bar\nu\bar\nu}_{ij,sh}f^{ij}_{E}(q^2)\vec{E}.
\end{align}
\end{subequations}
One can see that a magnetic field couples to both, the electric and the magnetic form factor. Also electric fields couple to both form factors. This can be understood as follows. In the neutrino rest frame, the magnetic field only couples to the magnetic dipole moment, and the electric field only couples to the electric dipole moment (if any), as suggested by the nomenclature. Lorentz covariance then demands that both electric and magnetic fields couple to the magnetic form factor in a system where the neutrino moves with nonzero velocity. A moving neutrino also exhibits spin precession in a pure electric field through its magnetic moment~\cite{Okun:1986uf}.

The Lorentz structure of Eqs.~\eqref{eq:nunumuB}--\eqref{eq:nunuepE} can now be readily calculated. In contrast to the previous sections, we neglect all contributions proportional to the mass since the magnetic and electric form factors are small and, in the models considered here, proportional to the neutrino mass already. The $(\gamma^0\vec{\gamma}\gamma_5)$ components are
\begin{subequations}
\begin{align}
\label{M1}
  \bigl(\gamma^0\vec{\gamma}\gamma_5\bigr)^{\nu\nu}_{ij,sh} &\approx  \left(\begin{array}{cc}0 & e^{+i\phi}\hat{\vec{\epsilon}}^{*}\\
  e^{-i\phi}\hat{\vec{\epsilon}}& 0\end{array} \right)\, ,\\
\label{M2}
  \bigl(\gamma^0\vec{\gamma}\gamma_5\bigr)^{\nu\bar\nu}_{ij,sh} &\approx \left(\begin{array}{cc}-e^{+i\phi}\hat{\vec{ p}} &0\\ 0& -e^{-i\phi}\hat{ \vec{p}}\end{array} \right)\,,\\
\label{M3}
  \bigl(\gamma^0\vec{\gamma}\gamma_5\bigr)^{\bar\nu\nu}_{ij,sh} &\approx  \left(\begin{array}{cc}-e^{-i\phi}\hat{ \vec{p}} & 0\\ 0& -e^{+i\phi}\hat{ \vec{p}}\end{array} \right)\,,\\
\label{M4}
  \bigl(\gamma^0\vec{\gamma}\gamma_5\bigr)^{\bar\nu\bar\nu}_{ij,sh} &\approx  \left(\begin{array}{cc}0 & -e^{-i\phi}\hat{\vec{\epsilon}}^{*}\\
   -e^{+i\phi}\hat{\vec{\epsilon}}& 0\end{array} \right)\,.
\end{align}
\end{subequations}

The remaining Lorentz structures are of the form $\left(i\gamma^0\vec{\gamma}\right)$. They read
\begin{subequations}
\begin{align}
\label{E1}
  \bigl(i\gamma^0\vec{\gamma}\bigr)^{\nu\nu}_{ij,sh} &\approx  \left(\begin{array}{cc}0& ie^{+i\phi} \hat{\vec{\epsilon}}^{*}\\
  -ie^{-i\phi} \hat{\vec{\epsilon}}&0\end{array} \right)\,,\\
\label{E2}
  \bigl(i\gamma^0\vec{\gamma}\bigr)^{\nu\bar\nu}_{ij,sh}&\approx  \left(\begin{array}{cc}-ie^{+i\phi}\hat{\vec{p}} & 0\\
  0& ie^{-i\phi}\hat{\vec{p}}\end{array} \right)\, ,\\
\label{E3}
  \bigl(i\gamma^0\vec{\gamma}\bigr)^{\bar\nu\nu}_{ij,sh}&\approx  \left(\begin{array}{cc}ie^{-i\phi}\hat{\vec{p}} & 0\\
  0& -ie^{+i\phi}\hat{ \vec{p}}\end{array} \right)\, ,\\
\label{E4}
  \bigl(i\gamma^0\vec{\gamma}\bigr)^{\bar\nu\bar\nu}_{ij,sh} &\approx  \left(\begin{array}{cc}0 & ie^{-i\phi} \hat{\vec{\epsilon}}^{*}\\
   -ie^{+i\phi} \hat{\vec{\epsilon}}& 0\end{array} \right)\, .
\end{align}
\end{subequations}

To this level of approximation, the $\nu\bar\nu$ and $\bar\nu\nu$ components are diagonal in helicity space, i.e.,  electric and magnetic fields mainly couple spin-0 neutrino-antineutrino pairs. Because the diagonal is proportional to $\hat{\vec{p}}$, the relevant field components are those parallel to the momentum of the neutrinos. The $\nu\nu$ and $\bar\nu\bar\nu$ components are off-diagonal in helicity space. The dominant effect of magnetic and electric fields on neutrinos and antineutrinos is spin precession. Here the transverse components of the electromagnetic fields contribute. The longitudinal components enter on the diagonals in the next order of the expansion in $m/E$ and are therefore omitted.

\subsection{Dipole moments of Majorana neutrinos \label{sec:MajMom}}

For Majorana neutrinos, electromagnetic transitions always contain two contributions, e.g.,
\begin{align}\label{eq:amp}
 \langle \nu_{\vec{p}_{\text{out}}}|\mathcal{H}^{\text{em}}|\nu_{\vec{p}_{\text{in}}}\rangle
 =A_\mu\left( \bar u_{\vec{p}_{\text{out}}}\Gamma^\mu u_{\vec{p}_\text{in}} - \bar v_{\vec{p}_{\text{in}}}\Gamma^\mu v_{\vec{p}_\text{out}}\right)\, .
\end{align}
This difference of two amplitudes leads to the cancellation of all the diagonal moments except for the anapole moment~\cite{Giunti:2014ixa}. This can also be understood by noting that the last two terms (including the minus sign) in Eq.~\eqref{eq:amp} are charge conjugates of each other. Because the Lorentz structure of the magnetic, electric, and charge form factors are \textit{C}-odd the combination vanishes. The Lorentz structure of the anapole moment is \textit{C}-even and does not cancel.

Because the magnetic moment of the Majorana neutrino vanishes, it does not couple directly to a magnetic field. However, magnetic fields polarize the background medium, and this effect does lead to helicity oscillations, see Sec.~\ref{sec:majorana}.

Electromagnetic moments of neutrinos depend on the details of the mechanism that creates the neutrino mass. When neglecting the model-dependent amplitudes, one can compare the moments of Dirac and Majorana neutrinos. The main differences are that the Majorana amplitudes contain Majorana PMNS matrices, which may contain more phases than Dirac PMNS matrices, and that Eq.~\eqref{eq:amp} has to be taken into account for Majorana neutrinos.

After these adjustments, the off-diagonal form factors of Majorana neutrinos can be obtained from Eq.~\eqref{eq:mmon1}. One finds that they depend on the relative \textit{CP}-phases of two neutrino species~\cite{Pal:1981rm}. The relative phase can either be equal or opposite, i.e., the ratio is $\pm 1$. For neutrinos with equal \textit{CP}-phases, the magnetic transition moments vanish~\cite{Pal:1981rm}, while for opposite \textit{CP}-phase the magnetic transition moments are nonzero and can be obtained from Eq.~\eqref{eq:mmon1} by substituting $\mathcal{F}_{ij}$ with  $2i\text{Im}\mathcal{F}_{ij}$.

For \emph{electric} dipole moments, the role of the \textit{CP}-phases is inverted. Opposite \textit{CP}-phases force the electric transition moments to vanish, while for equal \textit{CP}-phases the electric transition moments are nonzero and are obtained by substituting $\mathcal{F}_{ij}$ with $2\text{Re}\mathcal{F}_{ij}$~\cite{Pal:1981rm} in Eq.~\eqref{eq:mmon2}.

\subsection{Hamiltonian matrix for Majorana neutrinos}

The density matrix formalism naturally reproduces the results for the electromagnetic moments discussed in the last section. Similarly to Eq.~\eqref{eq:amp}, each component of the Hamiltonian matrix has two contributions from $\Gamma$ contractions, e.g.,
${\sf H}^{\nu\nu}_{ij,sh}(\vec{p})=\Gamma^{\nu\nu}_{ij,sh}(\vec{p})-\Gamma^{\bar\nu\bar\nu}_{ji,hs}(-\vec{p})$;
see Eq.~\eqref{GammasMajDef}.
The spinor contractions $\Gamma^{\nu\nu}$ and $\Gamma^{\bar\nu\bar\nu}$  have the same structure as for Dirac neutrinos;
see Eq.~\eqref{GammasDef}. Again neglecting the model dependence, the only difference is that the Dirac PMNS matrix have to be replaced by the Majorana PMNS matrix. For example, a magnetic field coupling to a Majorana neutrino through the magnetic form factor yields
\begin{equation}
\begin{aligned}
 {\sf H}^{\nu\nu}_{ij,\smi\spi} =& -\bigl[f^{ij}_{M}(q^2)-c.c.)\bigr]e^{+i\phi}\hat{\vec{\epsilon}}^{*}\vec{B}\\
 =&-2i\text{Im}[f^{ij}_{M}(q^2)]e^{+i\phi}\hat{\vec\epsilon}^{*}\vec{B}\, ,
   \end{aligned}
\end{equation}
where we have used the Hermiticity of the form factors. In the static limit, $2i\text{Im}[f^{ij}_{M}]$ is the magnetic moment of Majorana neutrinos~\cite{Shrock:1982sc}. It is zero for equal \textit{CP}-phases since $\mathcal{F}_{ij}$ becomes real. It is nonvanishing for opposite \textit{CP}-phases because $\mathcal{F}_{ij}$ becomes imaginary. An analogous argument holds for the electric dipole moment.

\section{\label{sec:helicity}Helicity coherence}

In this section we neglect  pair correlations and
discuss helicity coherence effects. To separate the latter from the
usual flavor coherence effects, we consider only one neutrino generation.
Furthermore, for definiteness we assume that neutrinos are Dirac particles.

\subsection{Order-of-magnitude estimate}

Two different mean-field backgrounds cause spin oscillations and create spin coherence: matter and neutrino currents, and electromagnetic fields. However, it is not clear which of these is dominant in a supernova. In the following we perform a crude estimate.

For Dirac neutrinos without pair correlations, the kinetic equations of neutrinos and antineutrinos decouple,
$i\dot{\rho}=[{\sf H}^{\nu\nu},\rho]$ and $i\dot{\bar\rho}=
[{\sf H}^{\bar\nu\bar\nu},\bar{\rho}]$, and, for one family, we only have to look at a $2\times2$ subsystem of the full evolution equation.
We start with a matter background with nonrelativistic velocity $\beta$, which flows orthogonal to the neutrino's momentum. The Hamiltonian matrix reads
\begin{align}
\label{GnunuMajExpl}
{\sf H}^{\nu\nu} &\approx
V\begin{pmatrix}
1 & \frac{m}{2p}\beta \\
\frac{m}{2p}\beta & 0
\end{pmatrix}\,,
\end{align}
where  $V$ is the usual  matter potential. For instance for $\nu_\mu$ or $\nu_\tau$
it is given by  \smash{$V=\GF n_n/\sqrt{2}$}, where
$n_n$ is the neutron density. We have omitted the neutrino kinetic
energy because it is diagonal in helicity space, and for a single generation
trivially cancels in the commutator.
The dependence 
of the diagonal terms of the Hamiltonian on the parallel flux and the 
dependence of the off-diagonal terms on the orthogonal flux were discussed 
in Refs.~\cite{Lobanov:2001ar,Studenikin:2004bu}.

In a derivation similar to the one that leads to Eq.~\eqref{GnunuMajExpl}, we obtain the $2\times2$
subsystem of the Hamiltonian matrix for a neutrino in a transverse magnetic field
\begin{align}
\label{GnunuMajExplMagn}
{\sf H}^{\nu\nu} &\approx
-\mu B
\begin{pmatrix}
0 & 1 \\
1 & 0
\end{pmatrix}\,
\end{align}
(see Sec.~\ref{sec:magnetic} for more details). Spin coherence is instigated by
the off-diagonals of Eqs.~\eqref{GnunuMajExpl} and~\eqref{GnunuMajExplMagn},
and to find the relative importance of the matter and magnetic contributions
it is sufficient to estimate their relative size. Typical magnetic fields in
a supernova are of order $10^{12} \,\rm{G}$ and much larger in magnetars.
Using the standard value for the magnetic moment given in Eq.~\eqref{eq:muD},
and assuming a neutrino mass of $0.1\, {\rm eV}$, we find for the contribution
of the magnetic field $\mu B \sim 10^{-16}\, {\rm eV}$. For a typical neutron
mass density $10^{12}~\rm{g/cm}^3$, which corresponds to a number density
$n_n\sim 10^4 \, \rm{MeV}^3$, the matter potential is of the order of the
neutrino mass, $V\sim 0.1\, {\rm eV}$. Thus, for a typical momentum $p\sim 30\, {\rm MeV}$
we obtain $V\beta m/(2p) \sim 10^{-10}\beta\, {\rm eV}$. For maximal background
velocities of $3000\, \rm{km/s}$, $\beta\sim 0.01$, the matter contribution
dominates. Surprisingly, the magnetic field is only important if the background
moves very slowly, if the matter density has decreased sufficiently, or if
the magnetic moment is enhanced.

Turning now to the density matrix, the size of the off-diagonal elements depends on the initial
conditions and history of the evolution. To obtain a rough estimate,  we can assume that the system has
reached equilibrium and, hence, its  previous evolution is irrelevant. In equilibrium,
the system is in an eigenstate of the Hamiltonian, i.e., $\mathsf{H}^{\nu\nu}$ and $\rho$
commute. This condition alone allows us to express the off-diagonals of the density matrix in
terms of the diagonals and components of the Hamiltonian matrix,
\begin{align}
\label{EqSolHelCoh}
\rho_{\smi\spi}&=\frac{\mathsf{H}^{\nu\nu}_{\smi\spi}}{\mathsf{H}^{\nu\nu}_{\smi\smi}-\mathsf{H}^{\nu\nu}_{\spi\spi}}
(\rho_{\smi\smi}-\rho_{\spi\spi})\,.
\end{align}
Keeping only the (dominant) matter contribution, Eq.~\eqref{GnunuMajExpl}, we find
$\rho_{\smi\spi}=(\rho_{\smi\smi}-\rho_{\spi\spi})\,m\beta/2p\sim m\beta/2p$. 
For $m\sim 0.1\, {\rm eV}$ and a typical momentum $p\sim 30\, {\rm MeV}$  this results in
$\rho_{\smi\spi}\sim 10^{-11}$, where we have used $\beta\sim 0.01$.

The same
result can be obtained by noting that if $\rho$ and $\mathsf{H}^{\nu\nu}$ commute, they can be
simultaneously diagonalized by a rotation that mixes positive- and negative-helicity states.
The rotation angle is $\tan 2\vartheta = m\beta/p$.  Considering e.g.\ the $\rho_{\smi\smi}=1$
eigenstate of the diagonalized Hamiltonian and rotating back to the basis where the Hamiltonian
has the form Eq.~\eqref{GnunuMajExpl} we find to leading order
\begin{align}
\label{rhoestimate}
 {\sf \rho}&\sim
 \begin{pmatrix}
 1 & \frac{m}{2p}\beta \\
 \frac{m}{2p}\beta & 0
 \end{pmatrix}
 \sim
 \begin{pmatrix}
 1 & 10^{-11} \\
 10^{-11} & 0
 \end{pmatrix}\,.
 \end{align}
The corrections to the diagonals are not included in  Eq.~\eqref{rhoestimate}
because they are of the order of $\delta \rho_{\smi\smi}\sim
\delta \bar\rho_{\spi\spi}\sim \rho^2_{\smi\spi} \sim 10^{-22}$
and are therefore negligibly small.

\subsection{Resonant enhancement}

For a magnetic field, the diagonal elements of the Hamiltonian matrix, Eq.~\eqref{GnunuMajExplMagn},
are zero for very relativistic neutrinos. This allows for the
magnetic fields to completely flip the spin of a population of neutrinos.
On the other hand, the diagonals of Eq.~\eqref{GnunuMajExpl} are in general nonzero
and suppress a complete conversion. In general, the matter contribution is given by
\begin{align}
\label{GnunuMajExplRel}
{\sf H}^{\nu\nu} &\approx
\begin{pmatrix}
V^0-V_{\shortparallel} & \frac{m}{2p}V_\perp \\
 \frac{m}{2p} V_\perp & 0
\end{pmatrix}\,,
\end{align}
see Eq.~\eqref{GammaMnunu},
where $V_{\shortparallel}\equiv \hat{\vec p}\vec{V}$ and
$V_{\perp}\equiv \hat{\vec \epsilon}\vec{V}$ are components of the matter flux
parallel and orthogonal to the neutrino momentum. Thus, if there are relativistic currents parallel to
the momentum of the neutrino such that the diagonals  vanish, Eq.~\eqref{EqSolHelCoh} implies
that a resonant enhancement of the spin conversion is possible.  
The possibility to generate the spin conversion 
by an orthogonal flux of matter, and the cancellation of the matter 
effect for relativistic matter moving along the direction of the neutrino momentum 
were first discussed in Refs.~\cite{Lobanov:2001ar,Studenikin:2004bu}
on the basis of the Lorentz-covariant Bergmann-Michel-Telegdi equation.
In Refs.~\cite{Vlasenko:2013fja,Vlasenko:2014bva} these effects have also 
been studied using the formalism of nonequilibrium quantum field theory.
In  the context of resonant leptogenesis the formation 
of flavor and helicity correlations in medium and the derivation of  
flavor-covariant transport equations able to account for helicity 
correlations has been discussed in Ref.~\cite{Dev:2014laa}.

For vanishing diagonals,
Eq.~\eqref{GnunuMajExplRel} can be rotated into its diagonal form with a rotation angle
$\vartheta=\pi/4$. In other words, mixing of the helicity states becomes maximal, similarly
to the Mikheyev-Smirnov-Wolfenstein resonance mixing, and hence in equilibrium
\begin{align}
{\sf \rho}&\sim
\begin{pmatrix}
\sfrac{1\!}2 & \sfrac{1\!}2 \\[1mm]
\sfrac{1\!}2 & \sfrac{1\!}2
\end{pmatrix}\,,
\end{align}
where we have again assumed that the system is in an eigenstate of the diagonalized
Hamiltonian. Outside of the core, a supernova is far from equilibrium,
but nonlinear feedback can enhance the spin-flipping processes~\cite{Vlasenko:2014bva}.

Making use of  Eq.~\eqref{SigmaDirac}, we can rewrite	
the resonance condition, $\mathsf{H}^{\nu\nu}_{\smi\smi}-\mathsf{H}^{\nu\nu}_{\spi\spi}=
V^0-V_{\shortparallel}=0$, in the form \cite{Vlasenko:2014bva}
\begin{align}
\label{SpinCoherenceResonance}
Y_e+\frac43\left(Y_\nu-\frac{V_{\shortparallel}}{2 n_b}\right)=\frac13\,,
\end{align}
where $Y_e\equiv n_e/n_B$ and $Y_\nu=(n_\nu-n_{\bar\nu})/n_B$ are the electron
and neutrino asymmetry fractions respectively and $n_B$ is the baryon number density,
The resonance condition can potentially be fulfilled  in or near the proto-neutron
star in a core-collapse supernova, or near the central region of a compact object
merger; see Ref.~\cite{Vlasenko:2014bva} and references therein.

\subsection{Lorentz covariance}

Helicity coherence builds up only if the
off-diagonal elements of the Hamiltonian matrix differ from zero. On the other hand,
because the off-diagonals are proportional to the component of the matter flow orthogonal
to the neutrino momentum, one can always find a frame where the off-diagonals
vanish and no helicity coherence builds up. In other words, at first
sight physical results seem to depend on the frame. This raises the question of
Lorentz covariance of the kinetic equations.

To be specific, let us consider the following simple example. We have two identical
observers moving with velocity $\beta$ with respect to each other. In the frame
of the first observer, the neutrino has momentum $\vec{p}$ along the $z$ axis and
the matter is at rest, i.e. $V_{\shortparallel}=V_\perp=0$,
\begin{align}
\label{Hframe1}
{\sf H}^{\nu\nu}&\approx V
\begin{pmatrix}
1 & 0 \\
0 & 0
\end{pmatrix}\,.
\end{align}
Thus no helicity coherence builds up. In the frame of
the second observer which moves with velocity $\beta$ along the $x$ axis the
Hamiltonian is no longer diagonal,
\begin{align}
\label{Hframe2}
{\sf H}^{\nu\nu}&\approx\frac{V}{\gamma}
\begin{pmatrix}
1 & \frac{m}{2p}\beta \\
\frac{m}{2p}\beta & 0
\end{pmatrix}\,,
\end{align}
and we expect helicity coherence to build up. Here $\gamma$ is the usual Lorentz
factor and $V$ and $p$ denote the potential and neutrino momentum in the frame
of the first observer.

Do the Hamiltonian matrices Eqs.\,\eqref{Hframe1} and \eqref{Hframe2} lead to different
physical results ? The answer is no, but to demonstrate this point we need to take
into account that a helicity state is also not Lorentz in\-variant. Let the neutrino
be in a state of definite helicity in the frame of the first observer, e.g.
$|p\hat{\vec z},\sm\rangle$, where $p$ is the absolute value of the neutrino
momentum and $\hat{\vec z}$ is the unit vector along the $z$ axis. The
corresponding density matrix reads
\begin{align}
\label{rhoframe1}
\rho=|p\hat{\vec z},\sm\rangle \langle p\hat{\vec z},\sm|=
\begin{pmatrix}
1 & 0 \\
0 & 0
\end{pmatrix}\,.
\end{align}
The Hamiltonian matrix Eq.~\eqref{Hframe1} and the density matrix commute and therefore the latter is
constant in time. The boost to the frame of the second observer transforms
$|p\hat{\vec z},\sm\rangle$ into a mixed helicity state with momentum
$\vec{q}$,
\begin{align}
|\psi\rangle = c_{\theta\hspace{-0.8pt}/\hspace{-0.6pt}2} |\vec{q},\sm\rangle -
s_{\theta\hspace{-0.8pt}/\hspace{-0.6pt}2} |\vec{q},+\rangle\,,
\end{align}
where $\theta$ is the angle of Wigner rotation around $\hat{\vec y}$ with
$\tan\theta=-m\beta/p$. Note that the rotation angle vanishes in the limit
of zero neutrino mass which reflects chirality conservation. The density
matrix develops off-diagonal elements,
\begin{align}
\label{rhoframe2}
\rho=|\psi\rangle\langle \psi|=\frac{1}{2}
\begin{pmatrix}
1+c_{\theta} & - s_\theta \\[1.5mm]
- s_\theta & 1-c_{\theta}
\end{pmatrix}
\,.
\end{align}
The Hamiltonian matrix and the density matrix again commute. In other words, the second
observer sees a mixed helicity state which, as expected, is also time
independent. This result reflects Lorentz covariance of the kinetic equations,
the lesson being that one has to transform the initial conditions consistently to obtain covariant results.

Let us now consider this result from a slightly different viewpoint. In each
frame, we can diagonalize the effective Hamiltonian by performing a Bogolyubov
transformation that mixes annihilation (creation) operators of the positive- and
negative-helicity states, $a_{s}\rightarrow c_\vartheta\,
a_{s}+s_\vartheta\, a_{-s}$. In particular Eq.~\eqref{Hframe2} is diagonalized
by a Bogolyubov transformation with the angle $\tan 2\vartheta=m\beta/p$. This
transformation brings the density matrix Eq.~\eqref{rhoframe2} back to the form
Eq.~\eqref{rhoframe1}. In other words, there is a connection between the Lorentz and
Bogolyubov transformations. In particular, if in every frame we diagonalize the
Hamiltonian then the transformed density matrix remains invariant under the boosts.

To summarize, as far as helicity coherence is concerned, both the Hamiltonian
and the density matrix transform under Lorentz boosts in such a way that the kinetic
equation is Lorentz covariant. We will rely on this result in
the discussion of particle-antiparticle coherence whose Lorentz transformation
properties are not as evident as for the helicity coherence.

\section{\label{sec:partantipart}Particle-antiparticle coherence}

In this section we discuss particle-antiparticle coherence.
In contrast to helicity coherence, which requires nonzero neutrino
masses, and flavor coherence, which in addition to nonzero masses
requires the existence of several neutrino generations, particle-antiparticle
coherence arises already for a single massless neutrino generation.
As has been discussed in the previous section, for a massless neutrino  the only
``natural'' correlators are $\rho_{\smi\smi}$, $\rho_{\spi\spi}$ and
$\kappa_{\smi\spi}$. To shorten the notation in this section we
suppress the spin indices. 

\subsection{Quantum-mechanical example}

To clarify the meaning of the particle-antiparticle coherence, let us first
study in more detail the simple quantum-mechanical example briefly discussed
in the Introduction.
We consider a system that can be in a linear combination of one of four pure states.
These are i) the empty state $|00\rangle$ without particles; ii) the paired
state $|11\rangle=a^\dagger(\vec{p}) b^\dagger(\sm\vec{p})|00\rangle$, which
contains a neutrino with momentum $\vec{p}$ and an antineutrino with momentum
$\sm\vec{p}$; iii) the one neutrino state $|10\rangle$; and  iv) the one antineutrino
state $|01\rangle$. Note that in all these states the antineutrinos
stream in the direction opposite to that of neutrinos. A general   state
can be expressed in terms of these four states,
$ |\psi\rangle= A_{00}|00\rangle+A_{11}|11\rangle+A_{10}|10\rangle+A_{01}|01\rangle$,
where the coefficients $A_{ij}$ are time dependent and normalized to unity,
$|A_{00}|^2+|A_{11}|^2+|A_{10}|^2+|A_{01}|^2=1$.

In analogy to Eq.~\eqref{HeffCrAn} we write the Hamiltonian in the form
\begin{align}
\label{twoModeH}
 H&=a^\dagger(\vec{p})\mathsf{H}^{\nu\nu} a(\vec{p}) +a^\dagger(\vec{p}) \mathsf{H}^{\nu\bar\nu} b^\dagger(\sm\vec{p})\nonumber\\
 &+b(\sm\vec{p}) \mathsf{H}^{\bar\nu\nu} a(\vec{p})- b^\dagger(\sm \vec{p})\mathsf{H}^{\bar\nu\bar\nu} b(\sm\vec{p})\,.
\end{align}
The Schr\"odinger equation for the coefficients $A_{ij}$ then splits into
three independent equations,
\begin{subequations}
\label{twoModeSchrod}
\begin{align}
 i\partial_t
 \begin{pmatrix}
 A_{00}\\
 A_{11}
 \end{pmatrix}
 &=\begin{pmatrix}
  0 & \mathsf{H}^{\bar\nu\nu}\\
  \mathsf{H}^{\nu\bar \nu}& \mathsf{H}^{\nu\nu}-\mathsf{H}^{\bar\nu\bar\nu}
  \end{pmatrix}
   \begin{pmatrix}
 A_{00}\\
 A_{11}
 \end{pmatrix}\,,\\
 i\partial_t A_{10} & = \mathsf{H}^{\nu\nu}A_{10}\,, \\
 i\partial_t A_{01} & = -\mathsf{H}^{\bar\nu\bar\nu}A_{01}\,.
\end{align}
\end{subequations}
Thus the evolution of the single-particle states completely
decouples because a homogeneous background medium cannot mix states of
different total momentum. On the other hand, the $|00\rangle$ and $|11\rangle$
states have zero momenta and therefore can be mixed by a homogeneous medium
through the $\mathsf{H}^{\bar\nu\nu}$ term of the Hamiltonian. However, the
$|00\rangle$ and $|11\rangle$ states have different angular momentum. Hence,
an anisotropic background medium, e.g. a transverse matter flux, is needed to
absorb the angular momentum and to mix the two states.

To make the connection to the  density matrix equations, we note that the
number of neutrinos and antineutrinos is given by $\rho = |A_{11}|^2+|A_{10}|^2$
and $\bar\rho= |A_{11}|^2+|A_{01}|^2$ respectively. Their time evolution can be
derived from Eq.~\eqref{twoModeSchrod} and takes the form expected from
Eq.~\eqref{RhoKappaEqWeyl},
\begin{subequations}
 \label{eq:Weylrho}
\begin{align}
 \dot \rho = -2  \text{Im}\left(\mathsf{H}^{\bar\nu\nu} \kappa\right)\, ,\\
 \label{eq:Weylrho2}
 \dot {\bar\rho}  = -2  \text{Im}\left(\mathsf{H}^{\bar\nu\nu} \kappa\right)\,,
\end{align}
\end{subequations}
if we identify $\kappa= A_{00}^*A_{11}$. Equation~\eqref{twoModeSchrod} also
leads to an evolution equation for $\kappa$
\begin{align}\label{eq:Weylka}
 i\dot \kappa & = \left(\mathsf{H}^{\nu\nu}-
 \mathsf{H}^{\bar\nu\bar\nu}\right)\kappa +
 \mathsf{H}^{\nu \bar \nu}\left(1-\rho -\bar \rho\right)\,,
 \end{align}
which can be obtained by using the normalization of the state $|\psi\rangle$.
Equation~\eqref{eq:Weylka} is again consistent with Eq.~\eqref{RhoKappaEqWeyl}
and coincides with the result of Ref.~\cite{Serreau:2014cfa} in the one-flavor
limit.

From these kinetic equations
we can infer that while $\rho$ and $\bar\rho$ are not separately conserved in
the presence of nonzero $\kappa$, their difference is conserved \cite{Serreau:2014cfa}.
Because $\rho(t,\vec{p})$ describes neutrinos with momentum $\vec{p}$ whereas $\bar\rho(t,\vec{p})$
describes antineutrinos with momentum $-\vec{p}$, the conservation of $\rho-\bar\rho$
implies that $\kappa$ induces the production
of neutrino-antineutrino pairs with opposite momentum.

The kinetic equation for $\kappa$
 describes a driven harmonic oscillator with frequency
$\mathsf{H}^{\nu\nu}-\mathsf{H}^{\bar \nu \bar \nu}\sim 2E$. Hence
$\kappa$ oscillates with twice the neutrino energy as expected.

From the definition $\kappa= A_{00}^*A_{11}$  we see that nonzero
parti\-cle-antiparticle coherence means that the system is not in an
eigenstate of the unperturbed Hamiltonian, but instead  in a mixture of
the $|00\rangle$ and $|11\rangle$ states, i.e., in a squeezed state.
Such states do not have a definite particle number. This observation clarifies the physical
meaning of particle-antiparticle coherence.

\subsection{Order-of-magnitude estimate}

As a next step we perform an order-of-magnitude estimate of $\kappa$.
For a single neutrino generation the extended density matrix
reduces to a $2{\times}2$ matrix of the form, see Eq.~\eqref{eq:simple-R-equation},  
\begin{align}
\label{Rmassless}
{\sf	 R}=
\begin{pmatrix}
\rho & \kappa\\
\kappa^\dagger & 1-\bar{\rho}
\end{pmatrix}\,,
\end{align}
where now $\rho$ and $\bar\rho$ are real numbers and $\kappa$ is a complex number.
We again start with an example of a matter
background with nonre\-la\-tivistic velocity $\beta$, which flows orthogonal
to the neutrino's momentum. Then, as follows from Eq.\,\eqref{GammaWeyl}, the
Hamiltonian matrix reads, see also Eq.~\eqref{eq:simplehamiltonian},
\begin{align}
\label{Hnonrel}
\mathsf{H}&=
E
\begin{pmatrix}
1 & 0 \\
0 & -1
\end{pmatrix}+
V
\begin{pmatrix}
1 & -\beta \\
-\beta & 1
\end{pmatrix}\,.
\end{align}
Unlike for helicity coherence, the neutrino kinetic energy $E$ no longer cancels out in the commutator.

Similarly to the case of spin coherence we can get a crude estimate of the $\kappa$
magnitude by assuming that the system has reached  equilibrium and hence $\dot \kappa=0$.
Equation~\eqref{eq:Weylka} then gives
\begin{align}
\label{kappaeqsol}
\kappa= -\frac{\mathsf{H}^{\nu\bar\nu}}{\mathsf{H}^{\nu\nu}-\mathsf{H}^{\bar\nu\bar\nu}}
(1-\rho-\bar{\rho})\,.
\end{align}
If we insert this result into Eq.~\eqref{eq:Weylrho} and use the Hermiticity of the
Hamiltonian matrix, we see that indeed $\dot \rho=\dot {\bar \rho} =0$.

Let us assume for a moment  that the neutrino-neutrino interactions are small
compared to the neutrino interaction with matter. For typical supernova parameters
$V\sim 0.1\,\rm{eV}$ and $E\sim 30\,\rm{MeV}$  and we then find $V/E\sim 10^{-9}$. Thus to a good
approximation $\mathsf{H}^{\nu\bar\nu}/(\mathsf{H}^{\nu\nu}-\mathsf{H}^{\bar\nu\bar\nu})
\sim V\beta/2E\sim 10^{-11}$, where we have used $\beta \sim 0.01$.  Because typically
$|1-\rho-\bar{\rho}|\sim 1$ we conclude that the ``natural''
size of the particle-antiparticle coherence is $\kappa\sim 10^{-11}$. 

The same result can be obtained by noting that in equilibrium $\sf R$ and $\sf H$ commute
and can be simultaneously diagonalized by a Bogolyubov transformation that mixes
neutrinos of momentum $\vec{p}$ with antineutrinos of momentum $-\vec{p}$. Under
this transformation the creation and annihilation operators transform as
\smash{$a(\vec{p})\rightarrow e^{-i\phi\hspace{-0.8pt}/\hspace{-0.6pt}2}c_\vartheta\,  a(\vec{p})+
e^{i\phi\hspace{-0.8pt}/\hspace{-0.6pt}2}s_\vartheta \,
b^\dagger(-\vec{p})$} and
\smash{$b^\dagger(-\vec{p})\rightarrow e^{i\phi\hspace{-0.8pt}/\hspace{-0.6pt}2}c_\vartheta\,
b^\dagger(-\vec{p})-e^{-i\phi\hspace{-0.8pt}/\hspace{-0.6pt}2}s_\vartheta\, a(\vec{p})$} res\-pe\-ctively,
where the phase $\phi=\arg {\sf H}^{\nu\bar\nu}$ and the rotation angle is given by $\tan 2\vartheta=2|{\sf H}^{\nu\bar\nu}|
/({\sf H}^{\nu\nu}-{\sf H}^{\bar\nu\bar\nu})\sim V\beta/E$. In the basis where the
Hamiltonian   is diagonal, the
system is described by (anti)neutrino densities, which we denote by $\varrho$ and $\bar\varrho$
respectively, and a pairing correlator, which we denote by $\varkappa$. From the transformation
properties of the creation/annihilation operators, we can infer the following relations
\begin{subequations}
\label{TwoBases}
\begin{align}
\rho&=c^2_\vartheta \varrho-c_\vartheta s_\vartheta \varkappa
-c_\vartheta s_\vartheta\varkappa^\dagger +s^2_\vartheta (1-\bar{\varrho})\,,\\
\bar\rho &=c^2_\vartheta \bar\varrho-c_\vartheta s_\vartheta \varkappa
-c_\vartheta s_\vartheta \varkappa^\dagger+s^2_\vartheta (1-\varrho)\,,\\
\label{kappaBogolyubov}
\kappa &=e^{i\phi}\left[c^2_\vartheta \varkappa+c_\vartheta s_\vartheta \varrho
-c_\vartheta s_\vartheta (1-\bar\varrho)-s^2_\vartheta \varkappa^\dagger\right]\,,
\end{align}
\end{subequations}
see Ref.~\cite{Vaananen:2013qja} for a detailed discussion.
Eigenstates of the diagonalized Hamiltonian are characterized by $\varkappa=0$.
Assuming, e.g., that the system is in an eigenstate of the diagonalized Hamiltonian
with some $\varrho$ and $\bar\varrho$, and rotating back to the basis where the
Hamiltonian has the form Eq.~\eqref{Hnonrel}, we find to leading order
\begin{align}
\label{Rexample}
 {\sf R}&\sim
 \begin{pmatrix}
 \varrho & \frac{V\beta}{2E}  \\[1mm]
  \frac{V\beta}{2E} & 1-\bar\varrho
 \end{pmatrix}\sim
 \begin{pmatrix}
 \varrho & 10^{-11}  \\[1mm]
  10^{-11} & 1-\bar\varrho
 \end{pmatrix}\,,
 \end{align}
which again leads to the tiny $\kappa\sim s_\vartheta\sim 10^{-11}$.

Pair correlations themselves are not measurable, and only their effect on the
number densities can be observed. A quick inspection of Eq.~\eqref{TwoBases}
shows that in equilibrium the difference between e.g. $\rho$ and $\varrho$ is of the order
of $s^2_\vartheta\sim \kappa^2$. In other words, the induced corrections to $\rho$
and $\bar \rho$ are quadratic in $\kappa$. 

This can also be understood from Eq.~\eqref{eq:Weylka}.
If the system has not yet reached equilibrium, then $\kappa$ oscillates around its
stationary value Eq.~\eqref{kappaeqsol}, provided that the components of the Hamiltonian
matrix only vary slowly with time compared to $\mathsf{H}^{\nu\nu} - \mathsf{H}^{\bar\nu\bar\nu}$. This assumption allows us to approximate the evolution of $\kappa$ as a driven harmonic oscillator with an amplitude that depends on the initial conditions. Assuming that pairing correlations are not created during neutrino production, the amplitude is of the order of the equilibrium value, Eq.~\eqref{kappaeqsol}.
We then find again that the mean number density created by pairing correlations is $\sim \kappa^2$.
Therefore,  the inclusion of the
particle-antiparticle coherence leads to a negligibly small $\delta \rho\sim
\delta \bar\rho \sim \kappa^2 \sim 10^{-22}$ .

\subsection{Including neutrino-neutrino interactions}

In the previous subsection we have estimated the ``natural'' size of $\kappa$ assuming that
the neutrino-neutrino
interactions are negligible. However, in a supernova the neutrino
density is very large and the neutrino background may play an important
role. This complicates the estimate of $\kappa$ because
$\mathsf{H}^{\nu\bar\nu}$ in Eq.~\eqref{kappaeqsol} itself depends on
$\kappa$ once we include neutrino-neutrino interactions,
\begin{equation}
\label{eq:selfnu}
\mathsf{H}^{\nu\bar\nu}= -V\beta-2\sqrt{2}\GF\,
\hat{\vec\epsilon}^{*}\!\!\int_{\vec{q}}[\hat{\vec{q}} \ell+\hat{\vec\epsilon}\kappa
+\hat{\vec\epsilon}^{*}\kappa^\dagger]\,,
\end{equation}
see Eqs.~\eqref{GammaWeyl} and \eqref{Inuvec}. A further complication
arises from the fact that the stationary value for $\kappa$ of one momentum mode
$\vec p$ depends on the pair correlations of all other momentum modes $\vec q$.
Note also   that the phase-space integral in Eq.~\eqref{eq:selfnu} is
unbounded. Pairing correlations with a momentum typical for the supernova
environment couple to pairing correlations of arbitrary high momentum. This
pushes us beyond the limitations of the Fermi approximation, and in principle
a fully renormalizable theory has to be studied to make sense of these 
high-momentum modes. To stay within the realm of applicability of the effective
theory,  we use a phenomenological cutoff $|\vec{q}|= M_W$ in the phase-space integrals.

To estimate the contribution of the $\kappa$ terms to the integral in
Eq.~\eqref{eq:selfnu}, we take into account that pair correlators of
different momentum modes oscillate incoherently such that we can
replace $\kappa$ by its approximate mean value,
$\kappa\approx-\mathsf{H}^{\nu\bar\nu}/2E$, where we use that $V\ll E$ and
assume $\rho+\bar\rho\ll 1$ in Eq.~\eqref{kappaeqsol}. To proceed we recall
that $\mathsf{H}^{\nu\bar\nu}=-\hat{\vec\epsilon}^{*}\vec{V}$, see
Eq.~\eqref{GammaWeyl}, where $\vec{V}$ is the total potential that includes
matter and neutrino contributions. Note further that $\vec V$
is momentum independent. With these substitutions, the integrals involving
$\kappa$ in Eq.~\eqref{eq:selfnu} read
\begin{equation}
\label{eq:cons1}
\text{Re}\int_{\vec{q}}\hat{\vec\epsilon}\kappa
\sim
\text{Re} \int_{\vec{q}}\hat{\vec\epsilon}\frac{\hat{\vec\epsilon}^{*}\!\cdot\!\vec{V}}{2E}
=\frac{\sqrt{2}}2\frac{\GF M_W^2}{3\pi^2}\vec{V}  \, ,
\end{equation}
where we have integrated up to the cutoff $|\vec{q}|= M_W$. Let us introduce the notation
\begin{equation}\label{eq:Hnol}
\mathsf{H}^{\nu\bar\nu}_0= -V\beta-2\sqrt{2}\GF\,
\hat{\vec\epsilon}^{*}\!\!\int_{\vec{q}} \hat{\vec{q}} \ell\,.
\end{equation}
Then using Eq.~\eqref{eq:cons1} we can write Eq.~\eqref{eq:selfnu} as
\begin{equation}
\mathsf{H}^{\nu\bar\nu}\approx\mathsf{H}^{\nu\bar\nu}_0\biggl(1-\sqrt{2}\frac{\GF M_W^2}{3\pi^2}\biggr)^{-1}\, .
\end{equation}
In other words the $\kappa$ terms in Eq.~\eqref{eq:selfnu} effectively lead to a renormalization of
the total potential produced by the matter and neutrino backgrounds. Numerically,  the correction is  small,
$\sqrt{2}\GF M_W^2/(3\pi^2)\approx 3\times 10^{-3}$, and can be   neglected.

In a supernova the neutrino density is comparable to that of matter.
Whereas each individual neutrino is relativistic, the bulk velocity of
the neutrino background is also comparable to the matter velocity. Thus,
the neutrino density contribution to Eq.~\eqref{eq:Hnol} is not expected to be larger than the matter
contribution. Furthermore, because the direction of the neutrino background
flux is more likely to be parallel to the momenta of individual neutrinos,
whereas the build up of the particle-antiparticle coherence requires a current
component orthogonal to the neutrino momentum, there is an additional
suppression as compared to the matter effect. All in all, the estimates
of $\kappa$ presented above remain essentially unaltered by the inclusion
of the neutrino-neutrino interactions.

\subsection{Resonance condition}

As follows from Eq.~\eqref{kappaeqsol}, particle-antiparticle coherence can
be resonantly enhanced if $\mathsf{H}^{\nu\nu}=\mathsf{H}^{\bar\nu\bar\nu}$.
In general for a relativistic matter flow  that also includes the neutrino
flux, the Hamiltonian matrix reads, see Eq.~\eqref{GammaWeyl},
\begin{align}
\mathsf{H}&=
E
\begin{pmatrix}
1 & 0 \\
0 & -1
\end{pmatrix}+
\begin{pmatrix}
V^0-V_{\shortparallel} & -V_\perp \\
-V_\perp & V^0+V_{\shortparallel}
\end{pmatrix}\,,
\end{align}
where, as before,  $V_{\shortparallel}\equiv \hat{\vec p}\vec{V}$ and
$V_{\perp}\equiv \hat{\vec \epsilon}\vec{V}$ are components of the matter flux
parallel and orthogonal to the neutrino momentum. The resonance condition then
translates into $E=V_{\shortparallel}$. Even assuming a relativistic matter flow,
for typical supernova parameters $V_{\shortparallel}/E\sim 10^{-9}$. In other
words, the resonance condition cannot be fulfilled in a supernova and there is
no reason to expect $\kappa$ to be larger than the estimate presented above.

Note also that for $V \sim E$ not only does the Fermi approximation break
down, but also the perturbative description is no longer applicable. In other
words it is in principle not possible to hit the resonance without rendering
the developed formalism meaningless.

\subsection{\label{sec:inconditions}Initial conditions}

All physical processes in which neutrinos are created  have time scales much larger
than the time scale of $\kappa$ oscillation. Hence, even during the production 
process neutrinos would adiabatically adapt to the propagation basis with respect to 
pair correlations. On the 
other hand, the time scales of flavor and helicity oscillation are much larger than
those associated with production and detection. This separation of time scales 
is crucial for the idea that neutrinos are produced in an eigenstate of interaction, i.e.,
in a coherent superposition of propagation eigenstates.
For the same physical reason, as neutrinos stream away from the supernova, they 
have enough time to adiabatically 
adapt to the external background. Thus, $\kappa$ does not oscillate but instead
closely tracks its equilibrium value. This makes dynamical equations for $\kappa$
essentially superfluous.  As the neutrinos
leave the supernova, the mean pair correlations approach zero adiabatically and
decouple from the evolution of $\rho$ and $\bar\rho$. 

\subsection{Lorentz covariance}

In the early Universe, the rest frame of the plasma is the only natural reference
frame and the question of Lorentz transformation properties of the pair
correlators does not arise \cite{Fidler:2011yq}. In a supernova environment the
situation is more complicated. In particular, the comoving frame of the matter
can in some cases be more convenient than the rest frame of a distant observer.
Similarly to helicity coherence, the particle-antiparticle coherence builds up
only if the off-diagonal components of the Hamiltonian matrix are not zero. However,
because the off-diagonals are proportional to the component of the matter flow
orthogonal to the neutrino momentum, their value depends on the frame. In particular,
one can   find a frame where the off-diagonals vanish and no particle-antiparticle
coherence builds up.  

Let us consider the same example as in Sec.~\ref{sec:helicity}. We have
two identical observers moving with velocity $\beta$ with respect to each other.
In the frame of the first observer, the neutrino has momentum $\vec{p}$ along
the $z$ axis and the matter is at rest, i.e. $V_{\shortparallel}=V_\perp=0$,
\begin{align}
\label{Hframe1pa}
\mathsf{H}&=E
\begin{pmatrix}
1 & 0 \\
0 & -1
\end{pmatrix}+
V
\begin{pmatrix}
1 & 0 \\
0 & 1
\end{pmatrix}
\,.
\end{align}
Thus no particle-antiparticle coherence builds up. In the frame of the second
observer which moves with velocity $\beta$ along the $x$ axis the Hamiltonian
is no longer diagonal,
\begin{align}
\label{Hframe2pa}
\mathsf{H}&=\gamma E
\begin{pmatrix}
1 & 0 \\
0 & -1
\end{pmatrix}+
V
\begin{pmatrix}
\sfrac{1\!}{\gamma} & -\beta \\
-\beta & \sfrac{1\!}{\gamma}
\end{pmatrix}
\,,
\end{align}
and we expect helicity coherence to build up. In other words, physical results
seem to depend on the frame.

As we have learned from the analysis of an analogous problem for helicity
coherence, the kinetic equations are covariant only if the initial conditions
also transform under the boost. Pair correlations ``couple''
neutrinos of opposite momenta. The notion of opposite momenta is not Lorentz
invariant and is violated by, e.g., a boost orthogonal to the neutrino momentum.
This alone implies that the initial conditions, which include specifying
$\kappa$ for all momentum modes, are not Lorentz invariant. At the same
time the very fact that the definition of $\kappa$ involves two momentum
modes makes it rather difficult to derive the corresponding Lorentz
transformation rules and we will not attempt the derivation here.

We have argued in the previous subsection that neutrinos are
produced and propagate in an eigenstate with respect to particle-antiparticle coherence. In
Sec.~\ref{sec:helicity} we have observed that if in every frame we
diagonalize the Hamiltonian then the (transformed) eigenstate of the
Hamiltonian remains invariant under the boosts. Here we assume that
the same holds true also for particle-antiparticle coherence. 
As a particularly interesting example let us assume that in the frame
of the first observer the system is in the vacuum state of the interacting
Hamiltonian, i.e. $\rho=\bar\rho=\kappa=0$,
\begin{align}
\label{Rframe1pa}
 {\sf R}&=
 \begin{pmatrix}
 0 & 0  \\
 0 & 1
 \end{pmatrix}\,.
\end{align}
The Hamiltonian matrix Eq.~\eqref{Hframe1pa} and the extended density
matrix Eq.~\eqref{Rframe1pa} commute and therefore the latter is constant
in time. According to our assumption, after diagonalizing Eq.~\eqref{Hframe2pa} 
by a Bogolyubov transformation, the transformed $\sf R$ takes the form  
Eq.~\eqref{Rframe1pa}. Transforming back to the initial basis we obtain,
\begin{align}
\label{Rframe2pa}
 {\sf R}&=
 \frac12
 \begin{pmatrix}
 1-c_\vartheta & s_\vartheta  \\
 s_\vartheta & 1+c_\vartheta
 \end{pmatrix}\,,
\end{align}
where $\vartheta$ is the angle of the Bogolyubov transformation that diagonalizes
Eq.~\eqref{Hframe2pa}, $\tan 2\vartheta=(\beta V/\gamma E)/[1-\beta^2(V/E)]$. By
construction the Hamiltonian matrix Eq.~\eqref{Hframe2pa} and the extended density
matrix Eq.~\eqref{Rframe2pa} commute and the latter is time inde\-pendent as well.

A perplexing feature of Eq.~\eqref{Rframe2pa}  is that it seems to describe a
state with a nonzero number of particles and antiparticles.
Whereas the first observer would see neither neutrinos nor antineutrinos, the
second observer that moves with respect to the first one with a \emph{constant}
velocity $\beta$ seems to observe a nonzero density of neutrinos and antineutrinos.
Put in other words, the empty space perceived by the first observer appears
to be filled with neutrino-antineutrino pairs in the frame of the second observer.
However, it is not entirely clear if the (anti)particle densities in Eq.~\eqref{Rframe2pa}
describe electroweak interaction eigenstates and thus would actually manifest themselves
via, e.g., particle production or momentum transfer to nuclei in scattering
processes. 

\subsection{Interpretation of the Bogolyubov transformation}

To better understand the meaning of the Bogolyubov transformation, 
we solve the equation of motion for a massless 
neutrino field coupled to a constant classical current $V^\mu$, and 
demonstrate that this solution reproduces the results obtained using 
the Bogolyubov transformation.  

In the Fermi limit 
\smash{$\mathcal{L}=\nu^\dagger_{\dot\alpha}\bar{\sigma}^{\mu,\dot{\alpha}\alpha}(i\partial_\mu-V_\mu)\nu_\alpha$}.	
Varying the Lagrangian with respect to the neutrino field, we obtain the equation of 
motion, $\bar{\sigma}^{\mu,\dot{\alpha}\alpha}(i\partial_\mu-V_\mu)\nu_\alpha=0$. 
Its solution can be written in a form similar to Eq.~\eqref{nudecompositionweil}, 
\begin{align}
\nu(t,\vec{p})&=a(t,\vec{p})\chi_{_-}(\hat{\vec{p}}_{\vec{V}})
+b^\dagger(t,-\vec{p})\chi_{_+}(\hat{\vec{p}}_{\vec{V}})\,,
\end{align}
where \smash{$a(t,\vec{p})=a(t,\vec{p})e^{-i\omega_+t}$} and 
\smash{$b^\dagger(t,-\vec{p})=b^\dagger(-\vec{p})e^{i\omega_-t}$} satisfy the usual anticommutation relations, 
$\hat{\vec{p}}_{\vec{V}}$ is the unit vector in the direction of 
$\vec{p}-\vec{V}$, and the energy spectrum is given by $\omega_\pm\equiv 
|\vec{p}-\vec{V}|\pm V^0$.

Using the orthogonal vectors $\hat{\vec p}$  and $\hat{\vec \epsilon}$ 
we can write $\omega_\pm$ in the form 
\smash{$\omega_\pm=\sqrt{(E-\hat{\vec{p}}\vec{V})^2+|\hat{\vec{\epsilon}}\vec{V}|^2}
\pm V^0$}, which reproduces the eigenvalues of the Hamiltonian matrix 
Eq.~\eqref{GammaWeyl}. The spinor contractions, see Eq.~\eqref{GammasDef},
now include $\chi_{_\mp}(\hat{\vec{p}}_{\vec{V}})$. By construction,
$\Gamma^{\nu\bar\nu}$ and $\Gamma^{\bar\nu\nu}$ vanish once we use the 
solution of the equations of motion. The diagonal elements can be
expanded  in terms of $\chi_{_\mp}(\hat{\vec{p}})$. For example for
$\Gamma^{\nu\nu}$ we obtain
\begin{align}
\label{chiexpansion}
\chi^\dagger_{_-}(\hat{\vec{p}}_{\vec{V}})
\bar{\sigma}^{\mu}\chi_{_-}(\hat{\vec{p}}_{\vec{V}})&=
c_1 n^\mu(\hat{\vec{p}})+\mathrm{Re}[c_2\epsilon^\mu(\hat{\vec{p}})]\,.
\end{align}
Multiplied by $V_\mu$, Eq.~\eqref{chiexpansion} reproduces the interaction
part of the $\mathsf{H}^{\nu\nu}$ element of the 
diagonalized Hamiltonian matrix. The decomposition coefficients 
\begin{align}
\label{c1and2}
c_1\equiv \frac{E-\hat{\vec{p}}\vec{V}}{|\vec{p}-\vec{V}|}\,\quad \mathrm{and} \quad
c_2\equiv -\frac{\hat{\vec{\epsilon}}^*\vec{V}}{|\vec{p}-\vec{V}|}\,, 
\end{align}
are related to the angle of the  Bogolyubov transformation by $c_1=\cos 2\vartheta$ 
and $|c_2|= \sin 2\vartheta$ respectively. 
In other words, diagonalizing the Hamiltonian matrix by a Bogolyubov transformation
in every frame is equivalent to using the equation of motion. This equivalence suggests interpreting
physical particle densities as propagation eigenstates of the full 
Hamiltonian in line with the discussion in Sec.~\ref{sec:inconditions}.

\section{\label{sec:summary}Summary and conclusions}

Neutrino flavor conversion is important in supernovae, yet a full
understanding remains elusive, largely because of neutrino-neutrino
refraction and concomitant self-induced flavor conversion, an effect caused
by run-away modes of the interacting neutrino gas. The difficulties in
developing a robust phenomenological understanding of even this relatively
simple case explains the reluctance to add further complications. Yet other
effects could be important as well, caused by inhomogeneities and
anisotropies of the medium and by magnetic fields, especially if one broadens
the view to include, for example, magnetars or neutron-star mergers. It is
often thought that helicity conversion effects will be small, at least if
neutrino dipole moments have no additional contributions beyond those
provided by their masses, yet one should remain open to such possibilities.
Finally, beyond flavor and helicity correlations, it has been stressed
recently that pair correlations could also become important.

Motivated by these concerns, we have studied extended kinetic equations that
describe flavor, helicity, and pair correlations, limiting ourselves to the
mean-field level, i.e., considering only propagation effects for freely
streaming neutrinos. Based on the ``forward Hamiltonian'' of neutrinos
interacting with a background medium, we have derived the various terms and
have given explicit results up to lowest order in the neutrino mass,  
similar to previous studies in the literature. For
Dirac neutrinos, we confirmed previous results and have extended them to 
include magnetic-field effects. For Majorana
neutrinos, we found a small correction to the mean-field Hamiltonian which
arises from lepton-number-violating contractions that appear only in the
Majorana case. To analyze the behavior of these additional terms in the limit
of vanishing neutrino masses, we have also studied extended kinetic equations
for Weyl neutrinos.

The density matrix formalism allows one to treat helicity oscillation induced
by matter currents  and by magnetic fields  on equal footing for both Dirac
and Majorana neutrinos. We have derived the mean-field Hamiltonian induced by
electromagnetic fields and compared it to that induced by matter currents.
Somewhat surprisingly, for typical supernova parameters, matter currents
dominate over magnetic fields. In principle, resonant enhancements can be
achieved, for example by relativistic flows of matter and background
neutrinos.

Flavor and helicity oscillations can be complicated in detail, but they are
conceptually straightforward. Their importance arises because charged-current
interactions produce neutrinos in flavor eigenstates, and all interactions
produce them in almost perfect helicity states. This nonequilibrium
distribution which is produced, for example, in the neutrino-sphere region of
a supernova, subsequently evolves coherently and leads to the various flavor
and helicity oscillation phenomena.

Concerning pair correlations, the mean-field equations produce similar
oscillation equations. In the simplest case of massless neutrinos, the pair
correlations are between neutrinos and antineutrinos of opposite momenta and
the oscillations are between the empty state and the one filled with a
neutrino and antineutrino. However, one probably cannot separate production
from subsequent propagation. The oscillation frequency is here twice the
neutrino energy, so in contrast to flavor and helicity oscillations, there is
no separation of scales between the energy of the state and the oscillation
frequency. Probably, as far as pair correlations are concerned, one should picture neutrinos as being produced in
eigenstates of propagation in the medium and not as eigenstates of the
interaction Hamiltonian. Flavor and helicity oscillations become important only
 because one produces a coherent superposition of different
propagation eigenstates. As this crucial characteristic appears to be missing
for pair correlations, we are tempted to suspect that pair correlations
remain a small correction to neutrino dispersion.

In the simplest case, helicity and pair correlations build up only in
anisotropic media because angular-momentum conservation forbids mixing of
states with different spin. If the anisotropy is a convective matter 
current, then there is a seeming paradox. In the frame with the current 
we expect correlations to build up. On the other hand, we may study 
these effects in the rest frame of the medium where no
correlations build up due to isotropy of the background. As far as 
helicity correlations are concerned, this  paradox is
resolved by noting that the handedness of massive neutrinos is not Lorentz
invariant. Transforming both the mean-field background and the neutrino
states to a different frame, e.g., the rest frame of the medium, leads to
consistent physical results. For pair correlations, physical results must
also be the same in all frames, yet it is less obvious   how to show this
point explicitly because the correlated modes of opposite momentum are
different ones in every frame.  Note, however, that in the
supernova context, there is not necessarily a natural coordinate system for
the study of neutrino propagation. Explicitly including production and
detection processes, i.e., the collision terms in the kinetic equation, may
shed more light on this question.

The ultimate ambition of fully understanding neutrino propagation in dense
environments and strong magnetic fields requires a more complete development
of its theoretical underpinnings. Our paper is meant as a contribution toward
this overall goal.

\section*{Acknowledgments}

We would like to thank I. Izaguirre, S. Chakraborty,  A. Dobrynina, and
C. Volpe for fruitful discussions.
We acknowledge partial support by the Deutsche Forschungsgemeinschaft (DFG)
under Grant No.\ EXC-153 (Excellence Cluster ``Universe''), and by the
Research Executive Agency (REA) of the European Union under Grant No.\
PITN-GA-2011-289442 (FP7 Initial Training Network ``Invisibles'').

\begin{appendix}

\section{\label{sec:spinorproducts}Chiral spinors}

Following the conventions of Ref.~\cite{Dreiner:2008tw}, which differ
from the ones used in Ref.~\cite{Serreau:2014cfa} by the overall sign of
$\gamma^0$, the Dirac matrices
in the Weyl representation, which is used in this work, are
\begin{align}
\gamma^\mu=\left(
\begin{tabular}{cc}
$0$ & $\sigma^\mu$\\
$\bar{\sigma}^{\mu}$ & $0$
\end{tabular}
\right)\,,
\end{align}
where $\sigma^\mu=(1,\vec{\sigma})$ and
$\bar{\sigma}^\mu=(1,-\vec{\sigma})$. Here $\vec\sigma$ is a
three-vector Pauli  matrix and $0$ and $1$ are $2{\times}2$
zero and unity  matrices respectively.
The chiral projectors are
\begin{align}
\label{PLPR}
P_L=
\begin{pmatrix}
1 & 0\\
0 & 0
\end{pmatrix}
\,,\quad
P_R=
\begin{pmatrix}
0 & 0\\
0 & 1
\end{pmatrix}\,.
\end{align}
The charge-conjugation matrix is
\begin{align}
\label{C}
C=-i\gamma^2\gamma^0=
\begin{pmatrix}
+\varepsilon & 0\\
0 & -\varepsilon
\end{pmatrix}\,,
\end{align}
where
\begin{equation}
\varepsilon=
\begin{pmatrix}
0&-1\\
+1&0
\end{pmatrix}\,.
\end{equation}
Notice that in two-component form, one usually writes
$\varepsilon_{\alpha\beta}$ for this antisymmetric $2{\times}2$ matrix and
\smash{$\varepsilon^{\dot\alpha\dot\beta}$} for $-\varepsilon$ appearing in
the lower right position of $C$.

In the Weyl representation and with these conventions, the Dirac bispinors are
\begin{subequations}
\label{ChiralAmplitudes}
\begin{align}
u_i(\vec{p},s)&=
\begin{pmatrix}
{\cal N}^i_{p,s}\,\chi_s(\hat{\vec p})\\[1mm]
{\cal N}^i_{p,-s}\,\chi_s(\hat{\vec p})
\end{pmatrix}
\,,\\
v_i(\vec{p},s)&=s
\begin{pmatrix}
-{\cal N}^i_{p,-s}\,\chi_{-s}(\hat{\vec p})\\[1mm]
{\cal N}^i_{p,s}\,\chi_{-s}(\hat{\vec p})
\end{pmatrix}
\,,
\end{align}
\end{subequations}
where $\hat{\vec p}$ is the unit vector in the direction of $\vec{p}$,
$p\equiv |\vec p|$, $s=\pm$ is a helicity index, and
\begin{align}
\label{Ndef}
{\cal N} ^i_{p,s}=\sqrt{\frac{E_{i}-sp}{2E_{i}}}\approx
\delta_{s-}+\frac{m_i}{2p}\delta_{s+}\,,
\end{align}
where $E_i=(p^2+m_i^2)^{1/2}$ is the energy of a neutrino with mass $m_i$.

We may describe the modes of the neutrino field in spherical
coordinates where the momentum components are
$\hat{\vec p}=(\sin\theta\cos\phi,\sin\theta\sin\phi,\cos\theta)$. In
this case, the standard two-component helicity spinors are explicitly
\begin{subequations}
\begin{align}
\chi_{_+}(\hat{\vec p})&=
\begin{pmatrix}
\cos\frac{\theta}{2}\\
e^{i\phi}\sin\frac{\theta}{2}
\end{pmatrix}\,,\\
\chi_{_-}(\hat{\vec p})&=
\begin{pmatrix}
-e^{-i\phi}\sin\frac{\theta}{2}\\
\cos\frac{\theta}{2}
\end{pmatrix}\,.
\end{align}
\end{subequations}
They satisfy the  orthogonality
condition $\chi^\dagger_{s}(\hat{\vec p})\chi_{h}(\hat{\vec p})=\delta_{sh}$.

The matrix elements of $\bar\sigma^\mu$ are then found by direct
evaluation to be
\begin{subequations}
\begin{align}
\chi^\dagger_{_-}(\hat{\vec p})\bar{\sigma}^\mu \chi_{_-}(\hat{\vec p})&=
n^\mu=(1,\hat{\vec p})\,,\\
\chi^\dagger_{_+}(\hat{\vec p})\bar{\sigma}^\mu \chi_{_+}(\hat{\vec p})&=
\bar n^\mu=(1,-\hat{\vec p})\,,\\
\chi^\dagger_{_+}(\hat{\vec p})\bar{\sigma}^\mu \chi_{_-}(\hat{\vec p})&=
-e^{-i\phi}\epsilon^\mu=-e^{-i\phi}(0,\hat{\vec \epsilon})\,,\\
\chi^\dagger_{_-}(\hat{\vec p})\bar{\sigma}^\mu \chi_{_+}(\hat{\vec p})&=
-e^{i\phi}\epsilon^{\mu*}=-e^{i\phi}(0,\hat{\vec \epsilon}^*)\,,
\end{align}
\end{subequations}
where $\epsilon^\mu$ is a polarization vector orthogonal to
$n^\mu$. The explicit components in spherical coordinates are
\begin{align}
\hat{\vec \epsilon}=
\begin{pmatrix}
e^{i\phi}\cos^2\frac{\theta}{2}-e^{-i\phi}\sin^2\frac{\theta}{2}\\[1mm]	
-i\bigl(e^{i\phi}\cos^2\frac{\theta}{2}+e^{-i\phi}\sin^2\frac{\theta}{2}\bigr)\\[1mm]
-\sin\theta
\end{pmatrix}
\,.
\end{align}
Note that the vectors $n^\mu$ and $\epsilon^\mu$ depend on $\hat{\vec p}$,
but we do not show this dependence explicitly to simplify the notation.

\section{\label{sec:contractions}Neutrino-neutrino mean-field Hamiltonian}

Because Majorana and Weyl neutrinos have two degrees of freedom, in many
cases it is more convenient to use two-component notation. For Majorana
neutrinos,
\begin{align}
\label{MajTwoComp}
\nu_i=
\begin{pmatrix}
\nu_{i,\alpha}\\
\nu_i^{\dagger\dot{\alpha}}
\end{pmatrix}
\quad\hbox{and}\quad
\bar{\nu}_i=\left(
\nu_i^\alpha\,\,\nu^\dagger_{i,\dot\alpha}
\right)\,,
\end{align}
where $\nu_{i,\alpha}$ and $\nu^\dagger_{i,\dot{\alpha}}$ are two-component
fields. They are related by Hermitian conjugation and transform under the
\smash{$\left(\frac12,0\right)$} and \smash{$\left(0,\frac12\right)$} representations of the
Lorentz group respectively. To emphasize the different transformation properties,
the conjugated fields, by convention, always carry a dotted spinor index.
The spinor indices $\alpha$ and $\dot{\alpha}$
are raised (lowered) using the spinor metric matrices \smash{$\varepsilon^{\alpha\beta}$}
and \smash{$\varepsilon^{\dot\alpha\dot\beta}$} (\smash{$\varepsilon_{\alpha\beta}$}
and \smash{$\varepsilon_{\dot\alpha\dot\beta}$}). Left-handed Weyl
fields satisfy the condition $P_L\nu=\nu$. Their explicit form  can be obtained from
Eq.~\eqref{MajTwoComp} by applying the chiral projectors.

Rewritten in terms of the two-component fields, the neutrino-neutrino
Hamiltonian density of Eq.~\eqref{Hself} is
\begin{align}
\mathcal{H}^{\nu\nu}&= \frac{\GF}{\sqrt{2}}\sum\limits_{ij}
\bigl[\nu^\dagger_{i,\dot{\alpha}}\bar{\sigma}^{\mu,\dot{\alpha}\alpha}\nu_{i,\alpha}\bigr]
\bigl[\nu^\dagger_{j,\dot{\beta}}\bar{\sigma}_\mu^{\dot{\beta}\beta}\nu_{j,\beta}\bigr]\,.
\end{align}
Taking expectation values of products of two of the four neutrino fields and bearing
in mind that fermions anticommute we obtain for the mean-field Hamiltonian
\begin{align}
\label{HeffTwoComponent}
\mathcal{H}^{\nu\nu}_{\rm mf}&= \frac{\GF}{\sqrt{2}}\sum\limits_{ij} \bar{\sigma}^{\mu,\dot{\alpha}\alpha}\bar{\sigma}_\mu^{\dot{\beta}\beta}\nonumber\\
&\times \Bigl[2\,\nu^\dagger_{i,\dot{\alpha}}\nu_{i,\alpha}\, \langle \nu^\dagger_{j,\dot{\beta}}\nu_{j,\beta} \rangle
-2\,\nu^\dagger_{i,\dot{\alpha}} \nu_{j,\beta}\, \langle \nu^\dagger_{j,\dot{\beta}} \nu_{i,\alpha}\rangle\nonumber\\
&\kern1.5em
+\nu^\dagger_{i,\dot{\alpha}} \nu^\dagger_{j,\dot{\beta}}\, \langle \nu_{j,\beta} \nu_{i,\alpha} \rangle
+ \nu_{i,\alpha} \nu_{j,\beta}\, \langle \nu^\dagger_{j,\dot{\beta}} \nu^\dagger_{i,\dot{\alpha}} \rangle\Bigr].
\end{align}

Translating back to the four-component notation we obtain
$[\bar{\nu}_i \gamma^\mu P_L \nu_i] \langle \bar{\nu}_j \gamma_\mu P_L \nu_j\rangle$
for the first term in Eq.~\eqref{HeffTwoComponent}. Using a Fierz identity \cite{Dreiner:2008tw},
$\bar{\sigma}^{\mu,\dot{\alpha}\alpha}\bar{\sigma}_\mu^{\dot{\beta}\beta}=
-\bar{\sigma}^{\mu,\dot{\alpha}\beta}\bar{\sigma}_\mu^{\dot{\beta}\alpha}$,
and translating back to the four-component notation we can represent the second term
in a similar form,
$[\bar{\nu}_i \gamma^\mu P_L \nu_j] \langle \bar{\nu}_j \gamma_\mu P_L \nu_i\rangle$.
By raising and lowering the spinor indices and reordering the fields the third
term can be rewritten as
$\sigma^{\mu}_{\dot{\alpha}\alpha}\bar{\sigma}_\mu^{\dot{\beta}\beta}
\nu^\dagger_{j,\dot{\beta}}\nu_{i}^{\dagger\,\dot{\alpha}}\,
\langle \nu_{i}^{\alpha} \nu_{j,\beta}\rangle$.
Using another Fierz identity  \cite{Dreiner:2008tw},
$\sigma^{\mu}_{\dot{\alpha}\alpha}\bar{\sigma}_\mu^{\dot{\beta}\beta}
=2\delta_\alpha\,^\beta\delta^{\dot \beta}\,_{\dot \alpha}$,
and translating back to  four-component notation we can rewrite
the third term in the form
$2[\bar{\nu}_j P_R C \bar{\nu}^T_i]\langle \nu^T_i C P_L \nu_j \rangle$.
Collecting all terms we obtain in four-component notation
\begin{align}
\mathcal{H}^{\nu\nu}_{\rm mf}=\sqrt{2}\GF\sum_{ij}\Bigl(&[\bar{\nu}_i \gamma^\mu P_L \nu_i] \langle \bar{\nu}_j \gamma_\mu P_L \nu_j\rangle\nonumber\\[-3mm]
&+ [\bar{\nu}_i \gamma^\mu P_L \nu_j] \langle \bar{\nu}_j \gamma_\mu P_L \nu_i\rangle\nonumber\\[4pt]
&+ [\bar{\nu}_i P_R  C \bar{\nu}^T_j]\langle \nu^T_j C P_L \nu_i \rangle \nonumber\\
&+ [\nu^T_i C P_L \nu_j ]\langle \bar{\nu}_j P_R C \bar{\nu}^T_i\rangle\Bigr)\,.
\end{align}
Using the definition of the charge-conjugate field, \smash{$\nu^c\equiv C\bar{\nu}^T$},
and the resulting $\overline{\nu^c}=\nu^T C$ we can further simplify  
and write the last two terms in a form which coincides with
Eq.~\eqref{Heff3},
\begin{align}
\mathcal{H}^{\nu\nu}_{\rm mf}=\sqrt{2}\GF\sum_{ij}\Bigl(&[\bar{\nu}_i \gamma^\mu P_L \nu_i] \langle \bar{\nu}_j \gamma_\mu P_L \nu_j\rangle\nonumber\\[-3mm]
&+ [\bar{\nu}_i \gamma^\mu P_L \nu_j] \langle \bar{\nu}_j \gamma_\mu P_L \nu_i\rangle\nonumber\\[4pt]
&+ [\bar{\nu}_i P_R \nu^c_j]\langle \overline{\nu^c_j} P_L \nu_i \rangle \nonumber\\
&+ [\overline{\nu^c_i} P_L \nu_j ]\langle \bar{\nu}_j P_R \nu^c_i\rangle\Bigr)\,.
\end{align}
Thus, the effective Hamiltonian obtained using the two-com\-ponent notation is identical
to the one obtained using the four-component notation, as expected.

\section{\label{sec:anapole}Right-chiral currents}

For completeness, we provide the contractions of the Lorentz structure of right-chiral currents $\left(\gamma^\mu P_R\right)$, which might arise in, e.g., beyond the Standard Model theories with right-handed currents. The contractions are
\begin{subequations}\label{GPR}
\begin{align}
(\gamma_\mu P_R)^{\nu\nu}_{ij,sh}  &\approx
\begin{pmatrix}
0 & e^{+i\phi}\frac{m_i}{2p}\epsilon^*_\mu \\
e^{-i\phi}\frac{m_j}{2p}\epsilon_\mu & n_\mu
\end{pmatrix}\,,\\[3pt]
(\gamma_\mu P_R)^{\nu\bar\nu}_{ij,sh}    &\approx
\begin{pmatrix}
e^{+i\phi}\frac{m_i}{2p}\bar n_\mu & 0 \\
\epsilon_\mu & e^{-i\phi}\frac{m_j}{2p}n_\mu
\end{pmatrix}
\,,\\[3pt]
(\gamma_\mu P_R)^{\bar\nu\nu}_{ij,sh}    &\approx
\begin{pmatrix}
e^{-i\phi}\frac{m_j}{2p}\bar n_\mu & \epsilon^*_\mu \\[2pt]
0 & e^{+i\phi}\frac{m_i}{2p}n_\mu
\end{pmatrix}
\,,\\[3pt]
(\gamma_\mu P_R)^{\bar\nu\bar\nu}_{ij,sh}&\approx
\begin{pmatrix}
\bar{n}_\mu & e^{-i\phi}\frac{m_j}{2p}\epsilon^*_\mu \\
e^{+i\phi}\frac{m_i}{2p}\epsilon_\mu & 0
\end{pmatrix}\,.
\end{align}
\end{subequations}
\end{appendix}
%

\end{document}